\documentclass[plaindraft,letter]{jpp-AAS}

\usepackage{graphicx,amsmath,amssymb}
\usepackage{natbib}
\usepackage[pdftex,colorlinks,citecolor=blue]{hyperref}
\usepackage{cleveref}
\usepackage{float}
\usepackage{subcaption}
\usepackage{enumitem}
\usepackage[percent]{overpic}

\def\apjs{Astrophys.~J.~Supp.~Ser.}
\def\apj{Astrophys.~J.}
\def\apjl{Astrophys.~J.~Lett.}

\def\mnras{Mon.~Not.~Roy.~Astron.~Soc.}

\def\prx{Phys.~Rev.~X}
\def\prl{Phys.~Rev.~Lett.}
\def\jgr{J.~Geophys.~Res.}

\def\prx{Phys.~Rev.~X}

\def\pop{Phys.~Plasmas}
\def\pof{Phys.~Fluids}

\def\pnas{Proc.~Nat.~Acad.~Sci.}

\def\jpp{J.~Plasma Phys.}
\def\jfm{J.~Fluid Mech.}

\def\sovast{J.~Soviet Astron.}
\def\njp{New J.~Phys.}
\ifCUPmtlplainloaded \else
  \checkfont{eurm10}
  \iffontfound
    \IfFileExists{upmath.sty}
      {\typeout{^^JFound AMS Euler Roman fonts on the system,
                   using the 'upmath' package.^^J}%
       \usepackage{upmath}}
      {\typeout{^^JFound AMS Euler Roman fonts on the system, but you
                   dont seem to have the}%
       \typeout{'upmath' package installed. JPP.cls can take advantage
                 of these fonts, if you use 'upmath' package.^^J}%
       \providecommand\upi{\pi}%
      }
  \else
    \providecommand\upi{\pi}%
  \fi
\fi

\ifCUPmtlplainloaded \else
  \checkfont{msam10}
  \iffontfound
    \IfFileExists{amssymb.sty}
      {\typeout{^^JFound AMS Symbol fonts on the system, using the
                'amssymb' package.^^J}%
       \usepackage{amssymb}%

      }{}
  \fi
\fi

\ifCUPmtlplainloaded \else
  \IfFileExists{amsbsy.sty}
    {\typeout{^^JFound the 'amsbsy' package on the system, using it.^^J}%
     \usepackage{amsbsy}}
    {\providecommand\boldsymbol[1]{\mbox{\boldmath $##1$}}}
\fi

\defcitealias{kswhite19}{Kunz, Squire {\it et al.} 2019}

\newcommand{\pD}[2]{\frac{\partial #2}{\partial #1}}
\newcommand{\D}[2]{\frac{{\rm d} #2}{{\rm d} #1}}
\newcommand\bb[1]{\mbox{\boldmath{$#1$}}}
\newcommand{\msb}[1]{\mathsfbi{#1}}

\newcommand{\imag}{{\rm i}}
\newcommand{\rmd}{{\rm d}}
\newcommand{\rme}{{\rm e}}
\renewcommand\bcdot{\,\bb{\cdot}\,}
\newcommand\btimes{\,\bb{\times}\,}
\newcommand\bdbldot{\,\bb{:}\,}
\newcommand\grad{\bb{\nabla}}

\newcommand{\eb}{\hat{\bb{b}}}
\newcommand{\vth}{v_{\rm th,i}}
\newcommand{\valf}{v_{\rm A}}
\newcommand{\vcs}{c_{\rm s}}

\title[High-$\beta$ collisionless wave interactions]{On hydromagnetic wave interactions in collisionless, high-$\bb\beta$ plasmas}

\author[S.~Majeski and M.~W.~Kunz]%
{S.~Majeski\ls$^{1,2}$%
  \thanks{Email address for correspondence: smajeski@princeton.edu}, 
M.~W.~Kunz$^{1,2}$}

\affiliation{$^1$Department of Astrophysical Sciences, Princeton University, Peyton Hall, Princeton, NJ 08544, USA\\[\affilskip]
$^2$Princeton Plasma Physics Laboratory, PO Box 451, Princeton, NJ 08543, USA\\[\affilskip]}

\pubyear{2023}
\volume{}
\pagerange{}
\date{\today}

\begin{document}

\maketitle

\begin{abstract}
We describe the interaction of parallel-propagating Alfv\'en waves with ion-acoustic waves and other Alfv\'en waves, in magnetized, high-$\beta$ collisionless plasmas. This is accomplished through a combination of analytical theory and numerical fluid simulations of the Chew-Goldberger-Low (CGL) magnetohydrodynamic (MHD) equations closed by Landau-fluid heat fluxes. An asymptotic ordering is employed to simplify the CGL-MHD equations and derive solutions for the deformation of an Alfv\'en wave that results from its interaction with the pressure anisotropy generated either by an ion-acoustic wave or another, larger-amplitude Alfv\'en wave. The difference in timescales of acoustic and Alfv\'enic fluctuations at high-$\beta$ means that interactions that are local in wavenumber space yield little modification to either mode within the time it takes the acoustic wave to Landau damp away. Instead, order-unity changes in the amplitude of Alfv\'enic fluctuations can result after interacting with frequency-matched acoustic waves. Additionally, we show that the propagation speed of an Alfv\'en-wave packet in an otherwise homogeneous background is a function of its self-generated pressure anisotropy. This allows for the eventual interaction of separate co-propagating Alfv\'en-wave packets of differing amplitudes. The results of the CGL-MHD simulations agree well with these predictions, suggesting that theoretical models relying on the interaction of these modes should be reconsidered in certain astrophysical environments. Applications of these results to weak Alfv\'enic turbulence and to the interaction between the compressive and Alfv\'enic cascades in strong, collisionless turbulence are also discussed.
\end{abstract}

%
%
\section{Introduction}\label{sec:intro}

Until recently, our understanding of Alfv\'en waves (AWs) at scales much larger than plasma-kinetic scales has been largely agnostic of the rate of Coulomb collisions. Indeed, linear shear AWs and nonlinear torsional AWs do not change the form of the velocity distribution function of the particles, but rather define the moving frame in which any changes to it are to be measured~\citep[e.g.,][]{schekochihin09}. As a result, it has been predicted that many defining characteristics of Alfv\'enic turbulence, such as its spectral indices or spatial anisotropies, are at most subtly affected by a lack of collisions. In introducing the concept of AW interruption however, \citet{squire16,squire17} established that shear AWs in collisionless, high-$\beta$ plasmas are susceptible to non-equilibrium effects that place  constraints on their ability to propagate freely, complicating the Alfv\'enic dynamics in new and novel ways.\footnote{$\beta \doteq 8\upi p/B^2$ is the ratio of the thermal pressure of the plasma particles, $p$, and the energy density stored in the magnetic field, $B^2/8\upi$, or equivalently twice the ratio of the squares of the sound speed $\vcs$ and the Alfv\'en speed $\valf$, {\em viz.}~$\beta\doteq 2\vcs^2/\valf^2$. In this paper, `high-$\beta$' means $\beta\gg 1$.} 
This is just one example of how even small deviations from local thermodynamic equilibrium (LTE), made possible by low collisionality,  can have large dynamical consequences for the evolution of high-$\beta$ systems. As a result, there has been increased interest in revisiting many problems in collisionless, high-$\beta$ plasmas that are fundamentally connected to Alfv\'enic dynamics, such as turbulence \citep{bott21,arzamasskiy23,squire23}, the fluctuation dynamo \citep{santoslima14,rincon15,sk18,St-Onge2020,sironi23,zhou23}, and magnetic reconnection \citep{cassak15,ak2019,wk19,egedal23}.

In well magnetized and collisionless plasmas, the importance of these non-equilibrium effects can be parameterized by the product $\beta \Delta$, where $\Delta \doteq p_\perp/p_\parallel -1$ is the normalized difference between thermal pressures across ($\perp$) and along ($\parallel$) the local magnetic-field direction. Aside from determining whether a plasma is unstable to various pressure-anisotropy-driven microinstabilities, the quantity $\beta \Delta$ also quantifies the competition between the plasma's {\color{black} pressure anisotropy} and the magnetic tension, with the characteristic propagation speed of magnetic disturbances being the effective Alfv\'en speed $v_{\rm A,eff} \doteq \valf\sqrt{1+\beta\Delta/2}$. When $\beta \Delta \sim \pm 1$, the restoring tension force can be significantly enhanced or even entirely eliminated, rendering Alfv\'enic fluctuations impotent. If $\beta \gtrsim 10$, then such circumstances are rather easy to achieve by even small deviations from LTE. 

The collisionless high-$\beta$ plasmas in which these effects appear are in fact pervasive throughout the universe. Indeed, from radiatively inefficient accretion flows onto supermassive black holes, to the hot and dilute intracluster medium of galaxy clusters, to certain regions of the solar wind at and beyond the Earth's orbit, it is often the case that Coulomb collisions are rare and magnetic fields, while strong enough to magnetize particles, are energetically weak \citepalias{kswhite19}. Further, turbulence is a ubiquitous phenomenon within these systems, and it plays a crucial role in processes like plasma heating \citep[e.g.,][]{sharma07,kunz18,kawazura19,arzamasskiy23} and angular-momentum transport \citep[e.g.,][]{sharma06,ksq16,bacchini22,sandoval23}. To  describe these processes within high-$\beta$ astrophysical environments accurately, some refinement of our understanding of their turbulent fluctuations is necessary. Therefore, with a focus on the role of $\beta\Delta$, this work revisits the following three tenets of Alfv\'enic turbulence that are known to hold in magnetohydrodynamics (MHD) and even $\beta\sim 1$ collisionless gyrokinetics: (i) that co-propagating AWs do not interact with one another, (ii) that weak interactions between AWs do not modify their field-aligned wavenumbers, and (iii) that  compressive fluctuations {\color{black} in the inertial range} are only passively mixed by the Alfv\'enic cascade\footnote{{\color{black}We are primarily concerned with effects on the inertial range of Alfv\'enic turbulence where the fluctuations are spatially anisotropic with respect to the background magnetic field with $k_\parallel/k_\perp\ll 1$ \citep[e.g.,][]{gs95}. Within intervals of the cascade not exhibiting this predominantly perpendicular polarization (e.g., near the forcing scale(s)), strong interactions between slow and Alfv\'en waves can occur \citep[e.g.,][]{hn13}}}.

The assertion that co-propagating AWs do not interact with each other in Alfv\'enic turbulence {\color{black}(point (i))} is a simple result of $v_{\rm A}$ being relatively insensitive to nonlinearities in $\beta \sim 1$ plasmas. This effect arises naturally from the MHD equations and governs wave interactions across both weak and strong turbulence. {\color{black}Specific to weak AW interactions however (point (ii)), one can effectively apply energy and momentum conservation to pairs of interacting waves since they remain correlated for many linear times.} Those constraints, combined with the fact that all Alfv\'enic fluctuations propagate at $v_{\rm A}$, prevent the generation of a cascade in $k_\|$. {\color{black}Lastly, for point (iii), inertial-range} compressive fluctuations being passively mixed by AWs results from the lack of overlap between the linearized eigenvectors of acoustic and Alfv\'enic fluctuations in $\beta\sim 1$ plasmas {\color{black}when $k_\perp \gg k_\|$}. None of the perturbed quantities of the acoustic waves enter to linear order in the propagation of AWs. However, among other things, AWs distort the direction of the magnetic field along which acoustic waves linearly perturb the thermal pressure. As such, they are sensitive to the presence of Alfv\'enic fluctuations. Yet, all of these phenomena may be expected to change in collisionless high-$\beta$ plasmas.

To understand the need for revisiting these wave interactions, consider the example of AW interruption \citep{squire16,squire17}. In the interruption process, a transverse perturbation ($\delta B_\perp$) to the magnetic field begins to oscillate at the Alfv\'en frequency and initially decreases in amplitude. As a result, the field strength $|B|\sim B_0 + \delta B_\perp^2/2B_0$ decreases nonlinearly. Conservation of the double adiabats $p_\perp/\rho B$ and $p_\parallel B^2/\rho^3$ then dictates that negative pressure anisotropy $\Delta<0$ is produced. If the amplitude is large enough ($\delta B_\perp/B_0 \gtrsim 2\beta^{-1/2}$), a pressure anisotropy of $\Delta \sim -2\beta^{-1}$ can be produced by the changing $|B|$, although limited to $\Delta = -2\beta^{-1}$ by the Alfv\'enic firehose instability threshold. However, the rate of anisotropy production is Alfv\'enic, while $p_{\perp/\|}$ are diffused by heat fluxes operating on sonic timescales ($\tau_{\rm sonic} \propto L /c_{\rm s} \ll L/v_{\rm A}$, where $c_{\rm s} \equiv \sqrt{(2p_\perp/3+p_\|/3)/\rho}$ is the sound speed). The anisotropy is then smoothed out by the heat fluxes and fills the domain with a nearly constant $\Delta = -2\beta^{-1}$. At this point, the magnetic tension that supports the AW is nullified and the fluctuation ceases propagation, hence the term `interruption'. \footnote{It has recently been found that certain conditions lead to the dominance of the resonant oblique firehose instability, which has a threshold of $\beta_\| \Delta \approx -1.4$ rather than ${\approx}-2$ (\citealt{hm00}; A.F.A.~Bott {\it et al.}, in preparation). At this more restrictive threshold, the magnetic tension is not nullified by the {\color{black} pressure anisotropy} and so full interruption does not occur. Within the purview of this work, we do not rely on the realizability of interruption, nor do we address the role of microinstabilities in detail.} If an AW is capable of experiencing a dramatic change in its propagation due to its own nonlinearly generated anisotropy, then one would expect that it would similarly respond to the anisotropy generated by other fluctuations within a turbulent system. It is for this reason that we revisit wave--wave interactions within high-$\beta$ collisionless plasmas.

In this paper, we focus on the dynamical and thermodynamical interactions between two waves: either ion acoustic waves (IAWs) and AWs, or two co-propagating  AWs. In \S\ref{sec:theory} we outline expectations for these interactions based off of analytical theory performed within the CGL-MHD framework with the help of simplified Landau-fluid heat fluxes \citep{cgl56,shd97}. Following this, in \S\ref{sec:sims} we test these predictions using a finite-volume Riemann-based code that numerically solves the CGL-MHD equations using the Landau-fluid heat fluxes. Finally, \S\ref{sec:disc} lays out the implications our results have for both weak and strong turbulence, and details some of the barriers that remain to establish a complete theory of Alfv\'enic turbulence in high-$\beta$ collisionless plasmas.

%
%
\section{Theory}\label{sec:theory}

The degrees of freedom associated with wave--wave interactions in a collisionless plasma are numerous, and so to  focus our investigation we make several simplifying assumptions. First, we take the electrons to be much colder than the ions, so that we may set their temperature $T_{\rm e} = 0$. Certain high-$\beta$ astrophysical plasmas are thought to produce these circumstances, such as radiatively inefficient accretion flows onto supermassive black holes \citep[e.g.,][]{quataert03}. However, the results we present are easily generalizeable to nonzero electron temperature (see~\S\ref{sec:disc}). Second, our focus is on the dynamics that occurs on wavelengths much larger than the ion Larmor radius and at frequencies much smaller than the ion gyrofrequency. Accordingly, we employ a fluid model that accounts for the adiabatic production of pressure anisotropy, its spatial redistribution by field-aligned heat fluxes, and its dynamical feedback on the plasma through the action of an anisotropic {\color{black} pressure}, all while neglecting finite-Larmor-radius effects. These `CGL-MHD' equations \citep{cgl56} are:
\begin{subequations}\label{eq:fullcgl}
\begin{equation}\label{eq:cont}
\pD{t}{\rho}+\grad \bcdot(\rho \bb{u})=0 ,
\end{equation}
\begin{equation}\label{eq:moment}
\rho\left(\pD{t}{\bb{u}}+\bb{u} \bcdot \grad \bb{u}\right)=-\grad\left(p_{\perp}+\frac{B^2}{8\pi}\right)+\grad \bcdot\left[\eb\eb\biggl(p_\perp - p_\parallel + \frac{B^2}{4\pi}\biggr)\right] ,
\end{equation}
\begin{equation}\label{eq:induc}
\pD{t}{\bb{B}}=\grad \btimes(\bb{u} \btimes \bb{B}), 
\end{equation}
\begin{equation}\label{eq:cglpprp}
\pD{t}{p_{\perp}}+\grad \bcdot\bigl(p_{\perp} \bb{u}\bigr)+p_{\perp} \grad \bcdot \bb{u}+\grad \bcdot\bigl(q_{\perp} \eb\bigr)+q_{\perp} \grad \bcdot \eb =p_{\perp} \eb \eb\bdbldot \grad \bb{u} ,
\end{equation}
\begin{equation}\label{eq:cglpprl}
\pD{t}{p_{\|}}+\grad \bcdot\bigl(p_{\|} \bb{u}\bigr)+\grad \bcdot\bigl(q_{\|} \eb\bigr)-2 q_{\perp} \grad \bcdot \eb =-2 p_{\|} \eb\eb\bdbldot\grad\bb{u}.
\end{equation}
\end{subequations}
Our notation is standard: $\rho$ is the mass density, $\bb{u}$ is the bulk fluid velocity, $\bb{B}$ is the magnetic field, and $p_\perp$ and $p_\parallel$ are the components of the pressure tensor oriented perpendicular and parallel to the local magnetic-field direction $\eb=\bb{B}/B$. Note that \eqref{eq:cont} and \eqref{eq:induc} may be used in \eqref{eq:cglpprp} and \eqref{eq:cglpprl} to obtain 
\begin{subequations}\label{eq:doubleadiabats}
    \begin{equation}
        p_\perp \D{t}{} \ln \frac{p_\perp}{\rho B} = - \grad\bcdot(q_\perp\eb) - q_\perp\grad\bcdot\eb ,
    \end{equation}
    \begin{equation}
        p_\parallel \D{t}{}\ln \frac{p_\parallel B^2}{\rho^3} = -\grad\bcdot(q_\parallel\eb) + 2q_\perp\grad\bcdot\eb  ,
    \end{equation}
\end{subequations}
where $\rmd/\rmd t\doteq \partial/\partial t + \bb{u}\bcdot\grad$ is the comoving time derivative. These equations express conservation of the first and second adiabatic invariants in the frame of the flow, but for the conservative redistribution of $p_\perp$ and $p_\parallel$ by the field-aligned conductive flows of perpendicular and parallel heat, $q_\perp$ and $q_\parallel$, respectively. Our third assumption is that these heat fluxes are given by the `3+1 model' of \citet{shd97}, which is tailored to capture the linear collisionless damping rates of small-amplitude fluctuations:
\begin{subequations}\label{eq:fullhf}
\begin{equation}
q_{\perp}=-\frac{2 c_{\rm s \|}^2}{\sqrt{2 \upi} c_{\rm s \|}\left|k_{\|}\right|}\left[\rho \nabla_{\|}\left(\frac{p_{\perp}}{\rho}\right)-p_{\perp}\left(1-\frac{p_{\perp}}{p_{\|}}\right) \frac{\nabla_{\|} B}{B}\right],
\end{equation}
\begin{equation}
q_{\|}=-\frac{8 c_{\rm s \|}^2}{\sqrt{8 \upi} c_{\rm s \|}\left|k_{\|}\right|} \rho \nabla_{\|}\left(\frac{p_{\|}}{\rho}\right),
\end{equation}
\end{subequations}
where $c_{\rm s,\|}=\sqrt{p_\|/\rho}$ is the parallel sound speed. The parameter $|k_\parallel|$ is a characteristic field-aligned wavenumber for the fluctuations in $B$, $p_{\perp/\|}$, and $\rho$ \citep[e.g.,][]{sharma06}; it is a simplification of the magnitude of the magnetic-field-aligned gradient operator, which would otherwise be very costly to calculate exactly because it is defined with respect to the local magnetic field and involves global operations. Because these heat fluxes describe linear collisionless damping, inherent in their use is the assumption that any perturbations produced by the waves studied here are sufficiently small with respect to the background plasma that nonlinear effects, such as saturation of the collisionless damping, can be ignored. Accordingly, while their analytical simplicity is invaluable in closing the CGL-MHD system, we discuss in \S\ref{sec:awaw} some circumstances in which these heat fluxes become inaccurate.

%
%
\subsection{Formulation of the problem}\label{sec:form}

Armed with \eqref{eq:fullcgl}, we now constrain the waves whose interactions are the focus of this paper. First, we consider only waves that propagate parallel to the background magnetic field $\bb{B}_0$, such that $k_\perp=0$. This choice removes the nonlinear Alfv\'enic interaction that normally drives a cascade to small scales, meaning that any interaction seen is solely a consequence of the effects of pressure anisotropy discussed in this work. Admittedly, many problems of interest involve fluctuations that possess $k_\perp/k_\parallel$ of order unity or even much greater, such as critically balanced Alfv\'enic turbulence \citep{gs95}. However, as we demonstrate in \S \ref{sec:iawaw}, one of the most important aspects of these wave interactions concerns the timescales associated with each interacting wave. Because AWs and IAWs possess frequencies that are independent of $k_\perp$, this facet of their interaction is captured regardless of $k_\perp$ {\color{black} (further discussion of the $k_\perp=0$ assumption can be found within \S \ref{sec:disc})}. Second, we assume that no constant background pressure anisotropy $\Delta_0$ is present in the plasma. A non-zero background pressure anisotropy may be included with little effect to the results (so long as the plasma remains stable) by replacing the Alfv\'en speed $v_{\rm A}\doteq B_0/\sqrt{4\upi\rho_0}$, where $\rho_0$ is the (assumed uniform) background density, with an effective Alfv\'en speed $v_{\rm A,eff}\doteq v_{\rm A}\sqrt{1+\beta\Delta_0/2}$.

Our chief interest in this paper is how an AW responds to the fluctuating pressure anisotropy that results from other waves -- a channel of communication that is customarily ignored in studies of plasma turbulence. Therefore, to distill this physics from \eqref{eq:fullcgl}, we pursue an asymptotic ordering of the wave perturbations that isolates the Alfv\'enic response to an externally generated pressure anisotropy $\Delta=\Delta(t,\bb{r})$. First, the AWs subject to $\Delta$ are taken to be small enough in amplitude that their own nonlinearities can be ignored on the timescales of interest here, {\em viz.}~${\delta \bb{B}_\perp/B_0\sim \bb{u}_\perp/v_{\rm A}\ll 1}$. With $k_\perp = 0$, the vectors $\delta \bb{B}_\perp$ and $\bb{u}_\perp$ can be oriented along any perpendicular coordinate without loss of generality, and so henceforth we refer only to $\delta B_\perp$ and $u_\perp$. Second, we take $\Delta$ to be large enough that its effect on the Alfv\'enic dynamics cannot be neglected, \textit{i.e.} $|\beta\Delta| \sim 1$. This ordering allows the {\color{black} stress} associated with the pressure anisotropy, $(p_\perp-p_\parallel)\eb\eb$, to be comparable to the Maxwell stress, $\bb{BB}/4\upi$. Lastly, we allow for compressive fluctuations in the parallel velocity ($u_\|$) and density ($\delta\rho$). Because $k_\perp = 0$ implies $\delta B_\parallel =0$, these fluctuations are purely acoustic, propagating at a speed ${\sim}c_{\rm s}$ that is much faster than $v_{\rm A}$. These assertions combined, the resulting maximal ordering is given by
\begin{equation}\label{eq:ord}
 |\Delta| \sim \frac{\delta B_\perp}{B_0} \sim \frac{u_\perp}{v_{\rm A}} \sim \frac{u_\parallel}{\vcs} \sim \frac{\delta\rho}{\rho_0} \sim \frac{1}{\beta}\doteq\epsilon\ll 1.
\end{equation}
To complete the ordering, we also make assumptions regarding the time evolution of these quantities. In particular, we assume that the time derivatives of $\delta B_\perp$ and $u_\perp$ are Alfv\'enic ($\partial_t \sim k_\parallel  v_{\rm A}$), while all other (compressive) quantities evolve on sonic timescales ($\partial_t \sim k_\parallel c_{\rm s}$). 

Without yet discussing the production/evolution of $\Delta$ (to be addressed in \S\ref{sec:iawaw} and \S\ref{sec:awaw}), we may then derive the evolution equations for our ordered Alfv\'enic fluctuations. Applying \eqref{eq:ord} to the components of \eqref{eq:moment} and \eqref{eq:induc} that are perpendicular to the guide field, and retaining only the leading-order terms in $\epsilon$, we find that
\begin{subequations}\label{eq:redalf}
\begin{equation}\label{eq:redalfind}
\pD{t}{\delta B_\perp} = B_0 \pD{x}{u_\perp},
\end{equation}
\begin{equation}
\rho_0 \pD{t}{u_\perp} = \frac{B_0}{4\upi} \pD{x}{B_\perp} + \frac{\beta}{2} B_0 \pD{x}{} \bigl(\delta B_\perp \Delta \bigr) .
\end{equation}
\end{subequations}
The transparency of these equations can be improved somewhat by rewriting them in terms of Els\"asser variables $z^\pm = u_\perp \pm \delta B_\perp /\sqrt{4\upi\rho_0}$ \citep{elsasser50}:
\begin{equation}\label{eq:elsass}
\pD{t}{z^\pm} \mp v_{\rm A} \pD{x}{z^\pm} = v_{\rm A} \frac{\beta}{4} \pD{x}{} \bigl[ (z^+ - z^-) \Delta \bigr] .
\end{equation}
The left-hand side of these Els\"asser equations describe forward ($-$) and backward ($+$) propagating AWs. The right-hand side provides the nonlinear interaction with another fluctuation that perturbs the pressure anisotropically. This anisotropy can arise from linear acoustic modes, or even other large-amplitude AWs that generate $\Delta$ nonlinearly. We analyze both of these situations in sections \ref{sec:iawaw} and \ref{sec:awaw} respectively.

%
%
\subsection{Interaction between ion-acoustic and Alfv\'en waves}\label{sec:iawaw}

In this section, we describe the interaction between monochromatic IAWs and AWs in an otherwise uniform plasma with density $\rho_0$ and pressure $p_0 = \beta B_0^2/8\pi$. This demands an understanding of the evolution of the $\Delta$ that appears on the right-hand side of \eqref{eq:elsass}. In our high-$\beta$ ordering \eqref{eq:ord}, the IAW fluctuations are energetically dominant over the Alfv\'enic fluctuations and, given their compressive nature, are chiefly responsible for the pressure perturbations $\delta p_\|$ and $\delta p_\perp$. This can be seen by applying \eqref{eq:ord} to the compressive components of \eqref{eq:fullcgl}, recalling that the time derivatives of compressive quantities are ordered as $\partial_t \sim k \vcs$, and retaining only leading-order terms:
\begin{subequations}\label{eq:lincomp}
\begin{equation}
\pD{t}{\delta \rho} + \rho_0 \pD{x}{u_\parallel} = 0 ,
\end{equation}
\begin{equation}
\rho_0 \pD{t}{u_\parallel} = -\pD{x}{\delta p_\parallel} ,
\end{equation}
\begin{equation}
\frac{1}{p_0} \pD{t}{\delta p_\perp} = \frac{1}{\rho_0} \pD{t}{\delta \rho} + \frac{\sqrt{2} c_{\rm s}}{\sqrt{\upi}|k_\parallel|p_0} \pD{x}{} \biggl( \pD{x}{\delta p_\perp} - \frac{p_0}{\rho_0} \pD{x}{\delta\rho}\biggr),
\end{equation}
\begin{equation}
\frac{1}{p_0} \pD{t}{\delta p_\parallel} = \frac{3}{\rho_0} \pD{t}{\delta\rho} + \frac{\sqrt{8} c_{\rm s}}{\sqrt{\upi}|k_\parallel| p_0} \pD{x}{} \biggl( \pD{x}{\delta p_\parallel} - \frac{p_0}{\rho_0} \pD{x}{\delta\rho} \biggr).
\end{equation}
\end{subequations}
These equations describing ion-acoustic fluctuations are entirely linear, with no feedback from the Alfv\'enic fluctuations. Therefore, we may simply prescribe the pressure anisotropy in \eqref{eq:elsass} being due to ion-acoustic fluctuations as 
\begin{equation}
    \Delta(t,x) = \Delta_0 \exp (\mathrm{i}k_{\rm C}x + \mathrm{i}\omega_{\rm C} t),
\end{equation}
where $\omega_{\rm C} \approx (\pm 1.47 + 0.46\mathrm{i})k_{\rm C} \vcs$ is the (complex) linear frequency of a damped ion-acoustic wave.\footnote{A non-decimal expression is available given that the `3+1' heat fluxes yield a cubic dispersion relation. However, the analytic expression in this case is more complicated than it is useful.} Substituting this anisotropy (with $+\omega_{\rm C}$ to start) into the Els\"{a}sser equations \eqref{eq:elsass} and Fourier transforming into $k$-space yields the following evolution equation for AWs at each wavenumber:
\begin{equation}\label{eq:ftelsass}
\frac{\partial z^\pm (k)}{\partial t} \mp \imag kv_{\rm A} z^\pm (k) = \imag kv_{\rm A} \frac{\beta\Delta_0}{4} \, \rme^{\imag\omega_{\rm C} t}  \bigl[ z^+(k-k_{\rm C}) - z^-(k-k_{\rm C}) \bigr].
\end{equation}
The right-hand side of \eqref{eq:ftelsass} shows that both forward- and backward-propagating 
AWs are generated at different wavenumbers when an AW is subject to the pressure anisotropy of an acoustic wave. Although this interaction term is not precisely proportional to the level of imbalance $|z^+(k)|^2-|z^-(k)|^2$, no change to either $z^+$ or $z^-$ will result if their initial amplitudes are exactly equal.

Assuming only an initial $z^-$, which makes the acoustic and Alfv\'en waves counter-propagating, the dynamics at the initial Alfv\'en mode's wavenumber $k_{\rm A}$ are simply 
\begin{equation}\label{eq:iawics}
 \frac{\partial z^\pm (k_{\rm A})}{\partial t} \mp \imag k_{\rm A} v_{\rm A} z^\pm (k_{\rm A}) = 0 \quad \Longrightarrow \quad z^- (k_{\rm A}) = z_0 \rme^{-\imag k_{\rm A}v_{\rm A}t};\; z^+(k_{\rm A})=0.
\end{equation}
The leading-order wave--wave interaction can be found at $k_{\rm A}+k_{\rm C}$ by evaluating \eqref{eq:ftelsass} given \eqref{eq:iawics}:
\begin{equation}\label{eq:collsln}
\begin{gathered}
z^+(k_{\rm A}+k_{\rm C}) \approx   z_0\frac{\beta\Delta_0}{4} \biggl( \frac{\omega_{\rm A}+k_{\rm C}\valf}{2\omega_{\rm A}-\omega_{\rm C}+k_{\rm C}\valf} \biggr) \left[\rme^{\imag(\omega_{\rm C}-\omega_{\rm A})t} - \rme^{\imag (k_{\rm A} +k_{\rm C})\valf t} \right], \\*
z^-(k_{\rm A}+k_{\rm C}) \approx  z_0\frac{\beta\Delta_0}{4} \biggl( \frac{\omega_{\rm A}+k_{\rm C}\valf}{\omega_{\rm C} +  k_{\rm C}\valf} \biggr) \left[\rme^{\imag(\omega_{\rm C}-\omega_{\rm A})t} - \rme^{-\imag (k_{\rm A} +k_{\rm C})\valf t} \right] ,
\end{gathered}
\end{equation}
where we have defined $\omega_{\rm A} \doteq k_{\rm A}v_{\rm A}$. Note that the choice $k_{\rm C} \sim k_{\rm A}$ implies $|\omega_{\rm C}| \gg \omega_{\rm A}$, and thus a weak interaction in high-$\beta$ plasmas. To obtain a stronger interaction, we focus on the amplitude of the reflected wave packet (for $\mathrm{Re}(\omega_{\rm C})>0$) and maximize the absolute value of the $z^+$ coefficient. Doing so indicates that the strongest interaction occurs when
 \begin{equation}\label{eq:kcka}
    \frac{k_{\rm C}}{k_{\rm A}} = \frac{2k_{\rm C}v_{\rm A}(k_{\rm C}v_{\rm A}+\omega_{\rm C,r})}{\omega^2_{\rm C,r}+\omega^2_{\rm C,i}-k^2_{\rm C}v^2_{\rm A}} \approx \frac{2}{\sqrt{\beta}},
 \end{equation}
where the subscripts `r' and `i' denote the real and imaginary parts of $\omega_{\rm C}$. With ${\beta \gg 1}$, equation~\eqref{eq:kcka} makes the absolute value of the $z^-$ coefficient ${\approx} 0.3z_0\beta\Delta_0$. For $\beta\Delta_0 \sim 1$, this is always an $\mathcal{O}(1)$ interaction, decreasing linearly with the amplitude of the acoustic mode. Perhaps unsurprisingly, the strongest interaction occurs when the frequencies are nearly matched, although this also means that there is effectively no change in wavenumber of the Alfv\'enic fluctuations, since $k_{\rm A}+k_{\rm C} \approx k_{\rm A} (1 + 2/\sqrt{\beta}) \sim k_{\rm A}$. In this sense, the strongest IAW--AW interaction resembles the frequency matching condition of the parametric decay instability \citep{sg69}. Additionally, while the maximum $z^+$ coefficient occurs for $\omega_{\rm C, r}>0$, the situation with $k_{\rm C} \sim k_{\rm A}/\sqrt{\beta}$ still yields an $\mathcal{O}(1)$ reflection if $\omega_{\rm C, r} <0$ (\textit{i.e.,} if the waves are co-propagating). 
 
In order to apply these conclusions to physical systems, we must also consider the case where $\Delta$ is purely real. This demands that the Fourier expansion of $\Delta$ contains both $+\imag k_{\rm C}$ and $-\imag k_{\rm C}$ exponentials, which in turn create source terms on the right-hand side of \eqref{eq:ftelsass} from both $k-k_{\rm C}$ \textit{and} $k+k_{\rm C}$. While this complicates the problem somewhat, equation~\eqref{eq:collsln} can still be used to obtain an order-of-magnitude estimate for the solution. The reason for this stems from the fact that the acoustic mode collisionlessly damps on the same timescale as it propagates. By the time an Alfv\'enic fluctuation at $k_{\rm A}+k_{\rm C}$ grows because of the interaction to a similar amplitude as that of the original wave at $k_{\rm A}$, $\Delta$ has largely vanished. The subsequent interaction of the $k_{\rm A}+k_{\rm C}$ AW with the now nearly depleted $\Delta$ to generate a $k_{\rm A}+2k_{\rm C}$ mode will be considerably weaker. As a result, the amplitude of the $k_{\rm A}+2k_{\rm C}$ fluctuation is likely to be small, and may not provide significant feedback to the evolution of the mode at $k_{\rm A}+k_{\rm C}$. This prediction is tested using numerical simulations in \S\ref{sec:sims}.

The conservation of certain quantities by this interaction may also be of interest. In particular, it may appear jarring that there is a nonlinear mechanism in \eqref{eq:elsass} by which new Alfv\'enic fluctuations can be generated, but no corresponding sink present in \eqref{eq:lincomp}. This would seem to imply that energy is not conserved if $|z^\pm|^2$ increase but nothing happens to the acoustic fluctuations. In this case, the apparent mismatch stems from the fact that, if $u_\parallel/\vcs \sim u_\perp/v_{\rm A}$, then there is considerably more (${\sim} \beta \times$) energy present in the acoustic fluctuations than in the Alfv\'enic ones. As a result, any energy given up by the acoustic fluctuations is higher order to themselves, but leading order to an AW with comparable amplitude. This can be demonstrated by going to third order in the compressive equations and re-establishing conservation of energy, a calculation presented in detail in Appendix~\ref{sec:appecons}.\footnote{For the same reason, the dynamical equations to leading order for each wave are unaffected by frequency matching. Although an increase in the gradients of Alfv\'enic quantities enhances their ability to affect the acoustic mode, the fact that {\color{black} those Alfv\'enic gradient terms} are multiple orders of $\epsilon$ smaller than the linear compressive terms means that increasing $k_{\rm A}$ by a factor of $\sqrt{\beta}$ or even $\beta$ will not render the solution \eqref{eq:collsln} inaccurate.} \textit{Why} energy goes to the Alfv\'enic from the acoustic fluctuations is not as obvious. The reason becomes apparent however, when considering the same interaction without damping of the acoustic wave. With $\omega_{\rm C,i}=0$ in \eqref{eq:collsln}, the leading-order Alfv\'enic response would then have a purely \textit{periodic} amplitude at each $k$. Essentially, the effects of a fluctuation with $\Delta>0$ would be nearly wiped out by the subsequent equivalent negative anisotropy. Only the \textit{root-mean-square} amplitude of the newly generated fluctuations would grow, and on a timescale much slower than the AW propagation. Instead, if the anisotropy decays, then the positive anisotropy generated during the first half of the acoustic period would be larger than the negative anisotropy generated during the second half-period, meaning that the difference in the nonlinear effects accumulates. Once the acoustic wave is decayed away, each $k$ of the Alfv\'enic spectrum is left with only a steady-state amplitude and energy.

%
%
\subsection{Interaction between two AWs}\label{sec:awaw}

The other source of pressure anisotropy we consider for \eqref{eq:elsass} is another, larger-amplitude AW. In particular, we are interested in the ability of nonlinear AWs to generate pressure anisotropy, thereby modifying their own $v_{\rm A,eff}$ in a manner that is dependent upon their amplitudes (similarly to the interruption process described in \S\ref{sec:intro}; \citealt{squire16}). We therefore choose to study the interaction between two co-propagating AW packets, which are initially isolated such that their self-modified $v_{\rm A,eff}$'s do not affect the other wavepacket. In this setup, one of the wave packets will be large enough in amplitude such that it generates a pressure anisotropy $\Delta \sim \beta^{-1}$, thereby fulfilling the role of the IAW in \S\ref{sec:iawaw}. This will be called the `primary' wave packet (with relevant quantities denoted by the subscript `p'). We first calculate the pressure anisotropy generated by this primary wave packet and demonstrate that its $v_{\rm A,eff}$ is increased as a result. Given sufficient time then, the two initially separated wave packets will be able to interact if the trailing wave packet is of larger amplitude. To study the subsequent interaction, we consider the leading (`secondary') wave packet to satisfy the Alfv\'enic components of \eqref{eq:ord}, thereby allowing us to employ \eqref{eq:elsass} in solving for its nonlinear deformation, while ignoring the back reaction on the primary AW packet.

The mechanism for producing pressure anisotropy differs considerably between IAWs and shear-Alfv\'en waves. While pressure anisotropy is produced at linear order for the former, the latter change neither the magnetic-field strength nor the plasma density to linear order in the fluctuation amplitude, and so by adiabatic invariance have no associated linear pressure anisotropy. Indeed, equations~\eqref{eq:doubleadiabats} without the heat fluxes give\footnote{Although ignored for illustrative purposes here, it is never justified to neglect the heat fluxes entirely in the evolution of high-$\beta$ collisionless plasmas. {\color{black}Owing to} their thermal nature the contribution of heat fluxes in the pressure evolution equations \eqref{eq:cglpprp} and \eqref{eq:cglpprl} is typically on the order of ${\sim}k_\parallel \vcs \delta p_{\perp/\parallel}$. In high-$\beta$ plasmas, this timescale ($k_\parallel \vcs$) is comparable to the rate-of-change of $p_{\perp/\parallel}$ generated by acoustic fluctuations ($\omega_{\rm C}$) and much larger than that of Alfv\'enic fluctuations ($k_\parallel \valf$).}
\begin{subequations}\label{eq:da}
\begin{equation}
    \D{t}{} \biggl(\frac{p_\perp}{\rho B}\biggr) = 0 \quad\Longrightarrow\quad \D{t}{p_\perp} = \frac{p_\perp}{\rho}\D{t}{\rho} + \frac{p_\perp}{B}\D{t}{B} ,
\end{equation}
\begin{equation}
    \D{t}{} \biggl(\frac{p_\parallel B^2}{\rho^3}\biggr) = 0 \quad\Longrightarrow\quad \D{t}{p_\parallel} = \frac{3p_\parallel}{\rho}\D{t}{\rho} - \frac{2p_\parallel}{B}\D{t}{B} .
\end{equation}
\end{subequations}
The magnetic-field strength changes only at second order in the fluctuation amplitude, according to $B = \sqrt{B_0^2 +\delta B_{\perp,\rm p}^2}$ (while $\delta \rho = 0$). From this, it is clear that $\delta B^2_{\perp,\rm p}/B^2_0$ must be ${\gtrsim}1/\beta$ in order for the {\color{black}stress} associated with the induced pressure anisotropy to be comparable to the Maxwell stress (the usual restoring force in an AW; \citealt{squire16}). In AW packets of this amplitude, the addition of a competitive {\color{black}pressure-anisotropic} stress dictates that their initial perturbation will be unable to propagate rigidly at $v_{\rm A}$ as is the case when $\Delta=0$. Instead, a strictly positive ({\color{black}owing to} $\delta B_{\perp, \rm p}^2 >0$) nonlinear pressure anisotropy will be generated by the oscillating $\delta B_{\perp, \rm p}$ with a dominant wavenumber of $k=2k_{\rm A}$. This cannot, however, last in the presence of the heat fluxes \eqref{eq:ord}. The pressure anisotropy $\Delta_{\rm p}$ is generated by the AW packet as it propagates at a rate of $\sim k\valf$, while it is diffused by the heat fluxes at a much faster rate of $\sim k\vcs$. As a result, $\Delta_{\rm p}$ is smeared out across the wave packet as it propagates, inheriting the spatial structure of the wave packet \textit{envelope} rather than the mode itself. The end result is that shear AW packets initialized with an amplitude $\delta B_\perp/B_0 \gtrsim \beta^{-1/2}$ generate smooth, envelope-scale, positive pressure anisotropies. Accordingly, $v_{\rm A,eff}>v_{\rm A}$ and such packets propagate faster than those with small (linear) amplitudes.\footnote{Note that we only address \textit{shear} AWs within this work. In the case of an infinite wave train of torsional AWs, no pressure anisotropy is produced, although they may still be affected by an externally supplied pressure anisotropy. In the case of a torsional AW packet, however, the front of the packet would still produce $\Delta > 0$ while the tail would generate $\Delta < 0$.}

With a $\Delta_{\rm p}$ profile resembling the envelope of the parent AW packet (rather than the oscillating wave within the envelope), there will be a peak in the magnitude of $\Delta_{\rm p}$ at the center of the packet, where $\delta B_{\perp, \rm p}^2$ is the largest. Through $v_{\rm A,eff} = v_{\rm A}\sqrt{1+\beta\Delta_{\rm p}(x)/2}$, this region of the packet will travel faster than the regions with lower $\delta B_{\perp, \rm p}^2$ that occur at the trailing edge of the envelope. In time, an isolated AW packet should therefore steepen to form a shock-like wavefront, which propagates at an enhanced $v_{\rm A,eff}$ from the background.\footnote{In \citet{squire16}, the authors also found that, as AW amplitudes approach the `interruption limit,' they are reshaped into square waves. Such a profile minimizes the variation in $|B|^2$, thereby minimizing the generation of $\Delta$. Although this effect has little impact on the interaction of the waves, it is observed in our simulations of large-amplitude AW packets.} It is this shocked, more rapidly propagating wave packet that can interact with smaller amplitude AW packets in front of it. In a collisionless plasma, however, the heat fluxes driven by a shock-like gradient in the pressure cannot be accurately described by \eqref{eq:fullhf}. This complicates our effort to solve \eqref{eq:cglpprp} and \eqref{eq:cglpprl} for the $\Delta_{\rm p}$ of the steepened packet. In this case, two alternative descriptions of $\bb{q}_{\perp/\|}$ are likely to be more accurate. For one, the heat fluxes may simply be that of free streaming particles. Unless reflected by the mirror force, plasma particles would transit the shock front and change their perpendicular temperature adiabatically according to $\delta T_{\perp, \rm p}/T_0 \approx \delta B_{\perp, \rm p}^2/2B_0^2$. Another possibility, which is likely for near-interruption AW packets in particular, is that the distribution of particle velocities resulting from an adiabatic response to $\delta B_{\perp, \rm p}$ would be unstable to kinetic microinstabilities. Larmor-scale magnetic fluctuations generated by the ion-acoustic or whistler heat-flux instabilities \citep{komarov18,roberg18} could scatter particles in pitch angle, giving the heat fluxes a similar functional form as \eqref{eq:fullhf}, but with a different effective diffusion rate. {\color{black} We choose to continue our analysis under this assumption of diffusive heat fluxes, both because it may apply in particular to near-interruption AW packets, and for the fact that our `3+1' Landau-fluid heat fluxes \eqref{eq:fullhf} are also diffusive, allowing us to verify the results numerically for arbitrary diffusion coefficients by modifying $|k_\||$ (appendix~\ref{sec:appawdp}). In this case,} given that particle motions are thermal, the upstream plasma with its higher temperature is able to pass through the shock front and diffuse downstream, at a rate controlled by the diffusion coefficient. This leads to the development of a pressure anisotropy precursor, which stretches in front of the shock as it propagates. For a finite diffusion coefficient, this precursor decays away steadily as one moves further downstream of the shock. We derive this precursor analytically in appendix~\ref{sec:appawdp} for a general diffusion coefficient $\kappa$, finding that the functional form of the pressure anisotropy is a decaying exponential in a frame moving with the shock.

\begin{figure}
    \centering
    \mbox{\hspace{3em}${\color{black}(a)}$\hspace{0.48\textwidth}${\color{black}(b)}$\hspace{0.4\textwidth}}\\
    \includegraphics[width=0.485\textwidth]{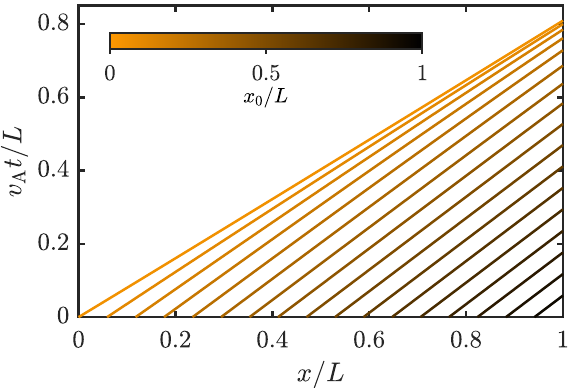}\quad
    \includegraphics[width=0.485\textwidth]{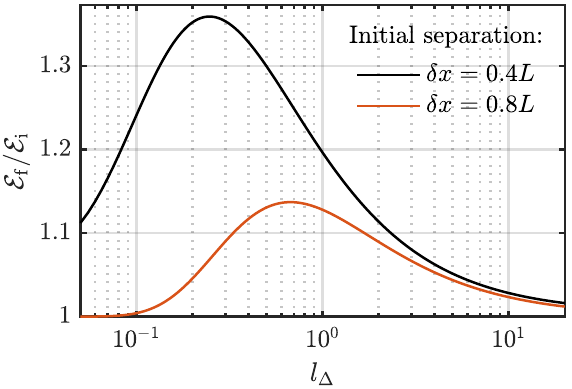}
    \caption{(a) Characteristic curves for an AW subjected to an exponentially decaying pressure anisotropy \eqref{eq:expDelta} with characteristic decay length $l_\Delta=0.1L$ and phase speed  $v_{\rm A,eff}=\sqrt{2}v_{\rm A}$. Characteristics originating near the left of the domain begin to converge by the time they reach the right side of the domain, indicating $\Delta$-induced AW steepening. (b) Energy gain by the secondary AW packet after one Alfv\'en-crossing time for different $l_\Delta$, demonstrating that a steeper $\Delta$ profile leads to more rapid gain in energy by the secondary packet.}
    \label{fig:AWtheory}
\end{figure}

What remains to be discussed is how the $\Delta_{\rm p}$ precursor of a large AW packet affects smaller-amplitude AW packets that cannot outrun its increased propagation speed, $v_{\rm A,eff}>v_{\rm A}$. As the tail of a secondary wave packet begins to feel the $\Delta_{\rm p}$ of the larger-amplitude primary packet approaching from behind it, there will be a larger $v_{\rm A,eff}$ at the back of the secondary wave packet than at the front. This will cause the secondary packet to compress, shortening its parallel wavelength. Unlike the IAW--AW interaction however, no backward propagating AW packet will be created. To see why, consider the evolution equation \eqref{eq:elsass} for $z^+$ assuming only an initial $z^-$:
\begin{equation}
    \frac{\mathrm{d} z^+}{\mathrm{d}t} \doteq \frac{\partial z^+}{\partial t} - v_{\rm A,eff}(x) \frac{\partial z^+}{\partial x}  = \valf\frac{\beta}{4} z^+ \frac{\partial \Delta_{\rm p}}{\partial x} - \valf\frac{\beta}{4} \frac{\partial}{\partial x} \bigl( z^- \Delta_{\rm p} \bigr).
\end{equation}
The final term on the right-hand side represents a possible source or sink, depending on its sign. However, for $\partial_x \Delta_{\rm p} <0$, the first term on the right-hand side damps $z^+$ exponentially along the wave characteristics. Given that the $\Delta$-precursor of a large-amplitude $z^-$ packet is monotonically decreasing in $x$, $z^+$ fluctuations are always damped in this interaction. This simplifies the equation for $z^-$, which can be written in terms of its energy density $\mathcal{E}^-(x) \doteq |z^-|^2$ by assuming $z^+=0$:
\begin{equation}\label{eq:lne}
    \pD{t}{} \ln{\mathcal{E}^-} + \valf \biggl(1+\frac{\beta\Delta_{\rm p}}{4} \biggr)\pD{x}{} \ln{\mathcal{E}^-} = -\valf\frac{\beta}{2}\pD{x}{\Delta_{\rm p}}.
\end{equation}
Solving \eqref{eq:lne} using the method of characteristics yields
\begin{equation}\label{eq:awawsln}
    \mathcal{E}^-(x,t) = \mathcal{E}_0^-(x_0(t,x)) \exp \biggl\{-\valf\frac{\beta}{2}\int\displaylimits_0^t\mathrm{d}t'\, \pD{x}{\Delta_{\rm p}}\biggr|_{x(t',x_0)}\biggr\},
\end{equation}
where $x(t,x_0)$ is the trajectory of the $x_0$ characteristic determined by the solution of
\begin{equation}
    \D{t}{x} = \valf\biggl(1+\frac{\beta\Delta_{\rm p}(t,x)}{4} \biggr),
\end{equation}
and $x_0(t,x)$ is the foot of the characteristic obtained by inverting $x(t,x_0)$. Equation~\eqref{eq:awawsln} states that the tail of the smaller-amplitude AW packet will steepen according to $\mathrm{d}_tx$, while growing exponentially along characteristics (and more rapidly so for steeper $\Delta_{\rm p}$). Motivated by the derivation of the $\Delta_{\rm p}$ precursor for diffusive heat fluxes in appendix~\ref{sec:appawdp}, we consider an exponentially decaying anisotropy
\begin{equation}\label{eq:expDelta}
    \Delta_{\rm p}(x) = \Delta_{\rm p, 0}\exp [(x-v_{\rm A,eff}t)/l_{\Delta}] ,
\end{equation}
which propagates rigidly at $v_{\rm A,eff} = v_{\rm A}\sqrt{1+\beta\Delta_{\rm p, 0}/2}$. Figure \ref{fig:AWtheory}(a) shows the characteristics that such an anisotropy would generate if $\Delta_{\rm p,0} = \beta^{-1}$ and $l=0.1L$, where $L$ is a characteristic domain length.
The compression of characteristics indicates that the trailing end of the secondary mode develops a shorter wavelength over time. Figure~\ref{fig:AWtheory}(b) exhibits the gain in energy experienced by a Gaussian wave packet that is subjected to this $\Delta_{\rm p}(x)$ after $t=L/v_{\rm A}$, for different values of the decay length $l_{\Delta}/L$. As $l_{\Delta}/L$ becomes smaller, the gain in energy of the mode begins to increase. In general this trend of increasing energy gain will continue, but here the gain curve drops when $l_{\Delta}/L$ becomes small enough that the packets becomes too well separated to feel the effect of $\Delta$ within a time $t=L/v_{\rm A}$ (for verification of this, we show the gain curve for two identical secondary packets initially separated from the primary by $\delta x = 0.4L$ and $\delta x = 0.8L$). Without the interference of other waves in this duo, the secondary packet compression will likely continue until the anisotropy of the primary wave damps away, or the wavelength of the secondary packet reaches dissipation scales. In a scenario where the large-amplitude wave interacts with many smaller-amplitude fluctuations, the compression (and therefore energization) of the smaller modes should be considered in the damping of the larger-amplitude mode, likely enhancing the rate of decay.\footnote{We have made no mention of energy conservation in the AW--AW interaction, even though the energy of the secondary packet increases. This is analogous to the IAW--AW interaction, given the $\sigma \sim \sqrt{\epsilon}$ ordering used in appendix~\ref{sec:appawdp}. The energy gain of the secondary packet is first order to itself, but higher order to the larger amplitude primary packet.}

We do not describe in any detail the interaction between primary and secondary anti-propagating AWs, however there is a notable asymmetry in the $\Delta$-mediated interactions between co- and anti-propagating AWs. Anti-propagating interactions are considerably weaker, both because of the much shorter interaction time caused by the larger difference in propagation velocity, and because of the difficulty of establishing any sort of frequency matching with an exponential pressure anisotropy. With co-propagating AW packets, the primary packet modifies $v_{\rm A,eff}$ nonlinearly, thus the difference in velocities between packets is proportional to $\delta B_\perp^2/B_0^2$; in the anti-propagating case, the difference in velocities is ${\approx}2v_{\rm A}$. Furthermore, due to the monotonically decreasing $v_{\rm A,eff}$ associated with the $\Delta$-precursor of the primary packet, the primary packet carries the co-propagating secondary packet with it and the two can interact until microphysical effects interfere. Alternatively, for an anti-propagating interaction, if one were to calculate the $k$-space dependence of the primary packet's $\Delta$, then the corresponding version of \eqref{eq:ftelsass} could be solved, but instead with numerous $k_{\rm C}$ because $\Delta$ would not be monochromatic. That said, unlike in the IAW--AW interactions, the fact that $\Delta$ is not monochromatic makes it difficult to achieve any semblance of frequency matching, and thus the interaction cannot similarly be tuned to maximize its strength.

%
%
\section{Numerical simulations}\label{sec:sims}

To test the predictions made in section~\ref{sec:theory}, we perform fluid simulations of each wave--wave interaction within the CGL-MHD framework \eqref{eq:fullcgl}. Our numerical approach utilizes a new Riemann solver implemented in a version of the finite-volume simulation code \texttt{Athena++} \citep{stone08} that evolves $p_\perp$ and $p_\parallel$ using CGL equations that are closed with Landau-fluid heat fluxes \eqref{eq:fullhf} \citep{squire23}. Unless otherwise stated, the field-aligned wavenumber $|k_\parallel|$ in the Landau-fluid heat fluxes is set equal to the initial dominant wavenumber of the primary pressure-anisotropy-generating fluctuation, {\em viz.}, the IAW in the acoustic--Alfv\'en interaction, and the larger-amplitude AW in the Alfv\'en--Alfv\'en interaction. All simulations are performed with $\beta=400$, with only linear perturbations being initialized for the fluid moments of each wave: $u_\perp$ and $\delta B_\perp$ for AWs; and $u_\parallel$, $\delta \rho$, and $\delta p_{\perp/\parallel}$ for acoustic waves. The interaction of acoustic and Alfv\'enic fluctuations is studied primarily using monochromatic plane waves, while Alfv\'en--Alfv\'en interactions are studied using wave packets to demonstrate how a larger-amplitude AW packet can `catch up' with an initially separated, smaller-amplitude AW packet. An acoustic--Alfv\'en interaction simulation was also performed using wave packets to visually depict the generation of a reflected AW as well as the non-locality of the interaction with respect to wavenumber. All monochromatic plane-wave simulations use periodic boundary conditions, while all simulations involving wave packets make use of outflow boundary conditions (where the derivatives of all fluid fields are set to zero at the boundaries).\footnote{In a periodic box, any finite-width wave packet that increases the local pressure anisotropy will, subject to heat fluxes, increase the background pressure anisotropy as well (and quite rapidly at high $\beta$). To simulate a wave packet that is much smaller than the extent of the background plasma then, only outflow boundary conditions can appropriately maintain the pressure anisotropy as a local perturbation, while removing the portion of $\Delta p$ that diffuses out away from the packet.} The only spatial coordinate is $x \in [0,1]$, aligned with the background magnetic field and measured in dimensionless Alfv\'enic units (where $v_{\rm A} = B_0/\sqrt{4\pi\rho_0}=1$) such that the Alfv\'en crossing time of a domain of length $L=1$ is $L/v_{\rm A}=1$. AW packets are initialized according to $\delta B_\perp = \alpha B_0 \psi(x)\cos (kx)$ and $u_\perp = \pm\alpha v_{\rm A} \psi(x)\cos (kx)$, with amplitude $\alpha$ and Gaussian envelopes $\psi(x)$ having standard deviations of approximately $2\upi/k$. {\color{black} For the AW--AW interaction, the primary packet amplitude is $\alpha = 4/\sqrt{\beta}$, while the secondary packet amplitude is $\alpha = 1/\beta$. In the IAW--AW interaction the acoustic wave is initialized with amplitude $\delta u_\parallel/v_{\rm th} = 2/3\beta$, and the Alfv\'en wave is initialized with $\delta B_\perp/B_0 = \delta u_\perp/v_{\rm A} = 1/2\beta$.} Because strong IAW--AW interactions at high $\beta$ are very non-local in $k$ space, AW fluctuations with wavelengths of as little as one-hundredth of the domain length need to be resolved. Similarly, for an AW packet to remain well separated from the outflow boundaries but still propagate for several linear times, the wave packet envelope and its dominant wavelength must be much smaller than the domain length. To  capture these high-$k$ fluctuations adequately, a resolution of 9,216 cells is used for the AW--AW interactions and up to 12,288 cells for the IAW--AW interactions. Importantly, the amplitudes of all fluctuations are chosen such that their associated pressure anisotropies are within the bounds of the firehose and mirror instabilities, $-2 \lesssim \beta \Delta \lesssim 1$ \citep{rudakov58,sk93}.

%
%
\subsection{Interaction between ion-acoustic and Alfv\'en waves}

\begin{figure}
    \centering
    \includegraphics[width=0.97\textwidth]{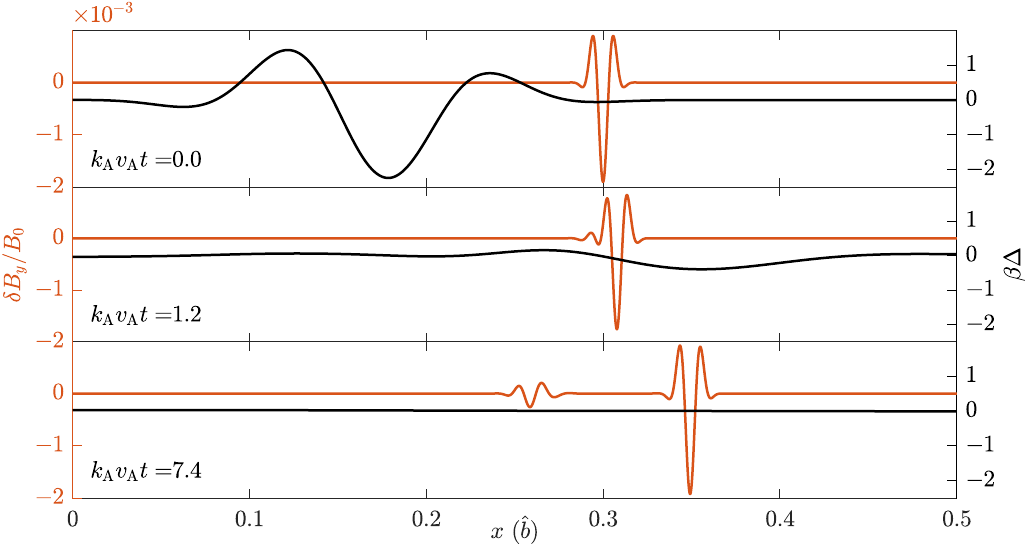}
    \caption{Landau-fluid CGL-MHD simulation of the interaction between the pressure anisotropy driven by an IAW packet (black) and an AW packet (orange) with matched frequencies. The IAW packet collisionlessly damps rapidly from the initial conditions (top frame), before nonlinearly deforming the short-wavelength AW packet (middle). After the IAW has decayed and propagated rightwards out of the domain, a backward-propagating AW packet has developed, with no apparent change to the wavenumber of the parent AW.}
    \label{fig:IAWevo}
\end{figure}

To illustrate the problem being studied, we begin with a simple example of co-propagating, quasi-monochromatic acoustic- and Alfv\'en-wave packets. We set the dominant wavenumber of the acoustic wave to $k_{\rm C} = k_{\rm A}/\sqrt{\beta}$, such that the modes are nearly  matched in their linear frequencies. Figure~\ref{fig:IAWevo} shows the evolution of these two packets from the initial setup until the point when the IAW has passed out of the domain, leaving behind only the modified AW packet(s). The top panel, showing the initial conditions, demonstrates that the initial pressure anisotropy associated with the IAW (black) dips as low as $\beta\Delta \approx -2$, and does not overlap with the AW to any significant degree. 
As the acoustic wave propagates, it decays rapidly at a rate of ${\simeq}0.31\omega_{\rm C}$. Therefore, by the time it overlaps with the AW, the amplitude of its driven pressure anisotropy has decayed significantly (middle panel). Nonetheless, the modification of the local $v_{\rm A,eff}$ can be seen in the tail of the AW as giving rise to a new, smaller-amplitude fluctuation. By the time the IAW has decayed away and propagated out of the simulation domain, two changes have occurred to the initial wave packet. Most obvious is that, as predicted by \eqref{eq:collsln}, a smaller-amplitude, backward-propagating AW packet of roughly equal wavelength to the parent AW packet has been generated. This process therefore appears to mimic some features of the parametric decay instability of AWs. It distinguishes itself from parametric decay, however, by the more subtle increase in the amplitude of the initial forward-propagating AW packet. Instead of reflecting a portion of the initial wave packet and decreasing its amplitude, the IAW spawns these modifications to the AW packet by giving up an infinitesimal (with respect to itself, not the AW) portion of its own energy. In the end, the degree of Alfv\'enic imbalance, normalized by the total energy of the AW fluctuations,  decreases by roughly $5\%$.

\begin{figure}
    \centering
    \includegraphics[width=0.96\textwidth]{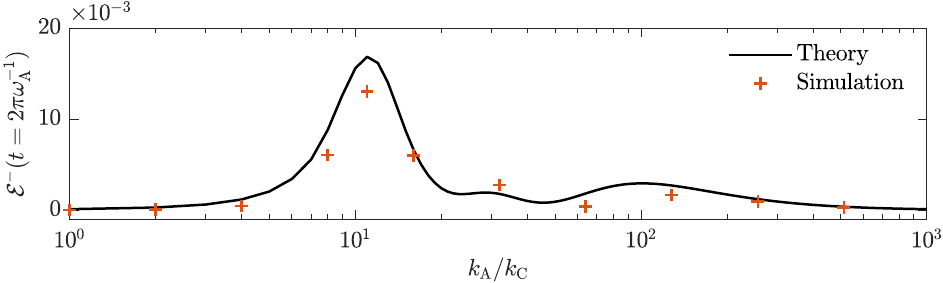}
    \caption{Energy contained within backward-propagating AWs after one Alfv\'en time of interaction between a forward-propagating monochromatic AW {\color{black} and a standing monochromatic IAW}, determined numerically as a function of the ratio of the Alfv\'en and acoustic wavenumbers using Landau-fluid simulations (red pluses). An approximate analytical solution for the energy \eqref{eq:oom} is shown as the solid line. Overall good agreement between theory and simulation is found, demonstrating strong interaction when the wave frequencies are approximately matched.}
    \label{fig:IAWoom}
\end{figure}

\begin{figure}
    \centering
    \mbox{\hspace{3em}${\color{black}(a)}$\hspace{0.44\textwidth}${\color{black}(b)}$\hspace{0.45\textwidth}}\\
    \includegraphics[width=0.48\textwidth]{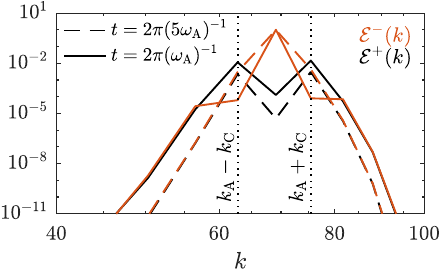}\label{fig:iawspec}\quad
    \includegraphics[width=0.47\textwidth]{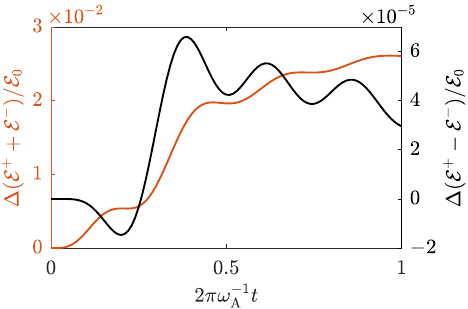}\label{fig:iawnrg}
    \caption{(a) Energy spectrum of forward- and backward-propagating Alfv\'enic fluctuations during interaction with a standing IAW. (b) Total change in wave energy (orange) and imbalance (black), {\color{black}normalized to the initial Alfv\'enic fluctuation energy $\mathcal{E}_0$,} versus time. In panel~(a), the IAW--AW interaction primarily generates new AW fluctuations at $k_{\rm A}\pm k_{\rm C}$, exhibited by the steepness of the spectra outside of the interval $k \in [k_{\rm A}-k_{\rm C}, k_{\rm A}+k_{\rm C}]$. This explains the accuracy of \eqref{eq:oom}, as fluctuations at $k_{\rm A} \pm 2k_{\rm C}$ are too weak to affect the solution dramatically. In panel~(b), the change in energy imbalance is several orders of magnitude smaller than the increase in total energy of the Alfv\'enic fluctuations, demonstrating that the IAW--AW interaction decreases imbalance with respect to overall AW energy.}
    \label{fig:spectra}
\end{figure}

The interaction between the wave packets in figure~\ref{fig:IAWevo} is expected to peak when the frequencies of the IAW and AW are roughly matched. However, the exact amplitude of the reflected fluctuation will depend on the minutia of the initial conditions, such as the packets' initial separation and the widths of their envelopes. We therefore move to the periodic plane-wave setup used to derive \eqref{eq:ftelsass}, so that we can assess the accuracy of \eqref{eq:collsln} with fewer variable factors. In doing so, we choose to initialize a standing IAW with zero initial pressure anisotropy (otherwise the intensity of the interaction would be hidden by an immediate Alfv\'enic response to the nonzero anisotropy at $t=0$). This is accomplished by perturbing $u_\parallel(x,t=0) = \alpha v_{\rm th}\sin (k_{\rm C}x)$ alone, with an amplitude {\color{black}$\alpha=2/3\beta$} such that the corresponding `forced' anisotropy in \eqref{eq:elsass} would be given by $\Delta(x) = \beta^{-1}\sin (\omega_{\rm C,r}t)\cos (k_{\rm C}x) \exp(-\omega_{\rm C,i}t)$. We simulate the interaction for a range of $k_{\rm A}/k_{\rm C}$, all for an elapsed time of $t_{\rm f}= 2\upi\omega_{\rm A}^{-1}$. The energy $\mathcal{E}^-=|z^-|^2$ contained within all $z^-$ fluctuations at the end of each simulation is plotted versus $k_{\rm A}/k_{\rm C}$ in figure~\ref{fig:IAWoom}, compared against an analytic estimate based off of \eqref{eq:collsln}.
The analytic estimate is simply obtained by accounting only for the initial AW within the source term on the right-hand side of \eqref{eq:ftelsass}. Separating the expression for the pressure anisotropy generated by the standing wave into its complex exponential parts, the approximation is given by
\begin{multline}\label{eq:oom}
    \mathcal{E}^-(t) \approx \frac{1}{8}\biggl[ \bigl|z^-(k_{\rm A}+k_{\rm C}, \omega_{\rm C,r}) - z^-(k_{\rm A}+k_{\rm C}, -\omega_{\rm C,r})\bigr|^2 \\*
    + \bigl|z^-(k_{\rm A}-k_{\rm C}, \omega_{\rm C,r}) - z^-(k_{\rm A}-k_{\rm C}, -\omega_{\rm C,r})\bigr|^2 \biggr],
\end{multline}
where the $\pm \omega_{\rm C}$ in parentheses indicates the direction of IAW propagation used when evaluating $z(k_{\rm A}+k_{\rm C})$. As expected, both the prediction and the analytic estimate peak in interaction strength when $k_{\rm A}/k_{\rm C} \sim \sqrt{\beta}/2\sim 10$, and agree rather well. The reason for this agreement can be found within figure~\ref{fig:spectra}(a), which shows the energy spectra of the forward- and backward-propagating Alfv\'enic fluctuations at $t=2\upi(5\omega_{\rm A})^{-1}$ and $t=2\upi(\omega_{\rm A})^{-1}$.
In figure~\ref{fig:spectra}(a), the slow, diffusive $k$-space nature of the IAW--AW interaction can be seen. With the initial AW represented by the peak in $\mathcal{E}^-(k)$, each Fourier component of the energy is several orders of magnitude smaller than the initial $k_{\rm A}$, with the sole exception being $k_{\rm A} \pm k_{\rm C}$ (denoted by two vertical dotted lines). Furthermore, the difference between $\mathcal{E}(k_{\rm A}\pm k_{\rm C})$ at $t=2\upi(5 \omega_{\rm A})^{-1}$ and $t=2\upi(\omega_{\rm A})^{-1}$ is much smaller than that for the higher/smaller $k$. Indeed, it is possible to show that, when energy cascades in both directions and $k_{\rm C}/k_{\rm A}$ is infinitesimal, equation~\eqref{eq:ftelsass} exhibits diffusive-like behaviour in $k$-space, with a diffusion rate that is  proportional to $k_{\rm C}^2$. Therefore, our estimate \eqref{eq:oom} is supported by this observation of $k_{\rm A}\pm k_{\rm C}$ being the dominant newly generated wavenumbers within a single Alfv\'en time. The $\mathcal{E}^\pm$ additionally evolve nearly identically, demonstrated by the relatively small change in cross-helicity with respect to the total energy shown in figure~\ref{fig:spectra}(b). This is expected, given that the form of \eqref{eq:ftelsass} suggests this interaction relates to the degree of imbalance at each wavenumber, and appears to effectively damp it over time.

%
%
\subsection{Interaction between two AWs}\label{sec:numawaw}

The interaction of co-propagating AWs relies upon the differing propagation speeds that result when different amplitudes generate different local pressure anisotropies. Accordingly, all simulations presented in this section will be of initially isolated wave packets with outflow boundary conditions. Both wave packets are initialized with a dominant wavenumber of $k_{\rm A} = 80\pi$, with amplitudes of $\delta B_z/B_0 = -\delta u_z/v_{\rm A} = 4/\sqrt{\beta}$ and $\delta B_y/B_0 = -\delta u_y/v_{\rm A} = \beta^{-1}$ for the large-amplitude and small-amplitude waves, respectively. The magnetic perturbations have been assigned to different directions only for the purpose of visualizing them separately; this choice is not necessary for the interaction itself. The standard deviations of the Gaussian wave envelopes are $2\pi/k_{\rm A}$ for both packets, the centres of which are initially separated by a distance of $\Delta x = 0.11$.
\begin{figure}
    \centering
    \mbox{\hspace{3em}${\color{black}(a)}$\hspace{0.45\textwidth}${\color{black}(b)}$\hspace{0.4\textwidth}}\\
    \includegraphics[width=1.0\textwidth]{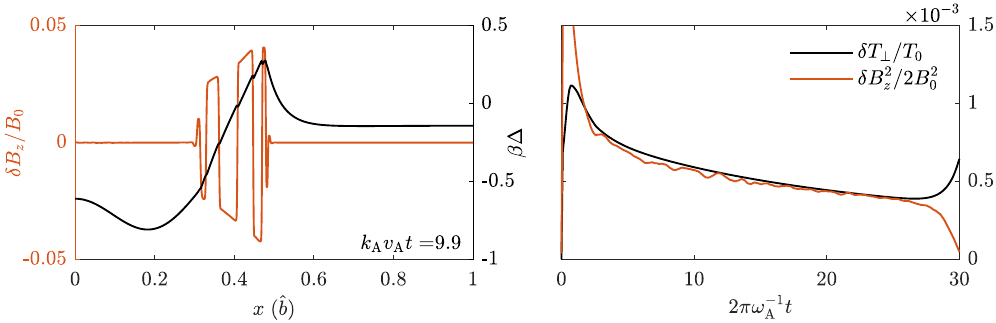}
    \caption{The magnetic field and pressure anisotropy profiles of the steepened AW packet (panel~a), alongside the relationship between the perpendicular pressure and the magnetic perturbation amplitude at the front of the shock (panel~b). The steepened wave packet is led by a shock in the magnetic-field profile, with a smoothed pressure anisotropy profile and $\Delta$ precursor modifying the local Alfv\'en speed. The dip in $\Delta$ behind the wave packet results from the initial conditions of the AW not including $\Delta$, hence the magnetic-field strength decreases in this region as the packet propagates away. As this packet propagates, the approximate conservation of the $T_\perp/B$ adiabat sets the amplitude of the decaying anisotropy, and thus $v_{\rm A,eff}$, at the shock front.}
    \label{fig:AWaniso}
\end{figure}

Before examining the interaction between these two wave packets, we first verify our predictions about the evolution of the primary wave packet. Figure~\ref{fig:AWaniso}(a) displays the primary AW packet after it has steepened to form a shock. As in \citet{squire16}, the waveform (orange) has become square-like in order to minimize the change in $\delta B_{\perp, \rm p}^2$. Due to the rapid action of the heat fluxes, the pressure anisotropy (black) has assumed a shape similar to that of the wave envelope, with an additional precursor that extends in front of the magnetic perturbation. A negative dip in $\beta\Delta$ is present behind the tail of the wave packet, although it does not propagate with the packet; it actually results from the magnetic perturbation departing from its initial location, causing a net decrease in $|B|$ at the location where it was initialized. {\color{black}This localized dip is akin to the negative anisotropy generated during the more conventional interruption process of a monochromatic AW \citep{squire17num}, as we did not initialize the wave packet with the pressure anisotropy that it generates shortly thereafter. Interruption does not occur here because the $\beta\Delta$ dip is not sufficiently negative to nullify the magnetic tension.} Figure~\ref{fig:AWaniso}(b) demonstrates the relationship between the perpendicular temperature and the perturbation to the magnetic-field strength. After an initial adjustment into the near-steady-state shocked AW packet, $\delta T_\perp$ closely follows the peak value of $\delta |B| = \delta B_{\perp \rm ,p}^2$ at the shock front. This supports our prediction made in appendix~\ref{sec:appawdp} that the maximum value of the pressure anisotropy can be calculated via conservation of the double adiabats.

\begin{figure}
    \centering
    \includegraphics[width=1.0\textwidth]{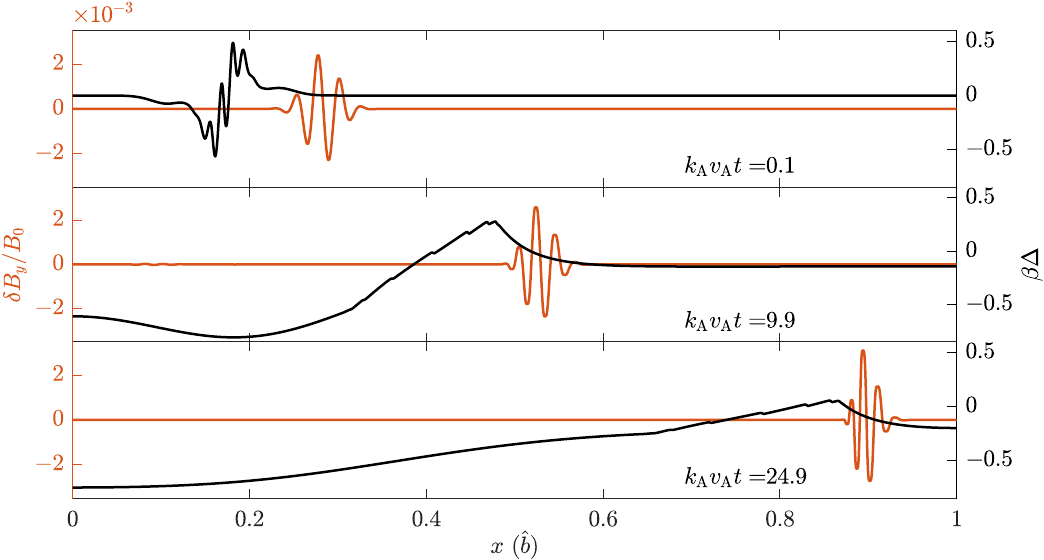}
    \caption{Time slices of a Landau-fluid simulation of the the AW-AW packet interaction. The pressure anisotropy of the primary packet is evolved alongside the magnetic perturbation of the secondary packet. The packets are initially separated (top), yet over time the enhanced effective speed allow the primary packet to approach the secondary from behind (middle). By the end of the simulation, the effective wavenumber of the secondary packet has more than doubled from the steepening induced by the primary packet's $\Delta$ precursor.}
    \label{fig:AWevo}
\end{figure}

In \S\ref{sec:awaw}, we predicted that the pressure-anisotropy-enhanced propagation speed of large-amplitude (primary) AW packets would allow them to `catch up' to and then distort smaller-amplitude (secondary) AW packets. Evidence of this physics can be found in figure~\ref{fig:AWevo}, which shows how the pressure anisotropy generated by the primary (black) interacts with the magnetic-field perturbation of the secondary (orange).
In the top panel, shortly after the magnetic disturbance of the large-amplitude mode begins to propagate, there is a short-wavelength $\Delta$. This pressure anisotropy is smoothed out rapidly by the action of heat fluxes, however, and leads to a smoothly increasing positive anisotropy perturbation whose front propagates at a slightly enhanced $v_{\rm A,eff}>v_{\rm A}$. The second panel at $k_{\rm A} v_{\rm A}t=9.9$ illustrates the moment that this front catches up to the tail end of the smaller-amplitude AW; the overlap between the two fluctuations causes this tail to propagate slightly faster, thereby compressing the waveform of the secondary. As this process continues into the final frame, the small-amplitude wave packet has had its width reduced by roughly half, and its amplitude increased due to the growth of $\mathcal{E}^-(x)$ (see \eqref{eq:awawsln}). Because the difference in Alfv\'en speeds between these two packets is relatively small, the cumulative deformation occurs slowly with respect to the Alfv\'en time of one of the packets, making this a weak interaction even though it leads to such a dramatic change in the structure of the wave packet envelope. This compression is expected to continue until the anisotropy of the large-amplitude packet has nonlinearly damped to be sufficiently small, or it is deformed by interaction with other wave packets.

%
%
\section{Discussion: applications in turbulence}\label{sec:disc}

The analysis presented in \S\S\ref{sec:theory} and  \ref{sec:sims} suggests that modern theories of turbulence in high-$\beta$ collisionless plasmas must account for fundamentally different wave--wave interactions. In weak MHD turbulence, energy and momentum conservation dictate that AWs are unable to modify their $k_\|$ through interaction \citep{mt81,galtier2000}. Yet, we have just demonstrated a weak interaction of two AWs with differing amplitudes where the parallel wavelength does in fact change significantly. In a similar vein, the compressive cascade of MHD turbulence relies on the fact that slow modes possess a linear timescale that never exceeds the Alfv\'enic timescale of the turbulence that mixes them. Further, these slow modes have no effect on the Alfv\'enic cascade and effectively decouple from it below the forcing scale \citep{lithwick01,cho2003,schekochihin09,nazarenko11,hn13}. However, when slow modes are replaced by collisionless IAWs, these roles nearly reverse, with acoustic modes propagating too fast to be  mixed effectively by the AWs, and in some cases even demonstrating an ability to reflect Alfv\'enic fluctuations.

Still, certain aspects of these waves that may be important in a turbulence setting are not considered in this study. In particular, because we focused solely on wave amplitudes for which the induced pressure anisotropies satisfy $|\Delta|\lesssim 1/\beta$, we do not consider the effects of the firehose or mirror instabilities. Previous work has shown that AWs, IAWs, and other compressive fluctuations evolve quite differently if their self-generated pressure anisotropies are large enough in magnitude to excite these Larmor-scale instabilities \citep{squire17num,kunz20,mks23}. As demonstrated from first principles by \citet{arzamasskiy23}, the anomalous scattering that results from these instabilities is likely to influence the entire turbulent cascade, yielding yet another source of non-local energy transfer. 

As mentioned in \S\ref{sec:theory}, this study was performed  assuming cold electrons. However, generalizing to electrons with finite temperature is quite straightforward, as long as the electrons remain barotropic. With an electron pressure tensor that satisfies $\msb{p}_{\rm e} = p_{\rm e} \msb{I} = (\rho T_{\rm e} /m_{\rm i})\msb{I}$, no additional pressure anisotropy is produced by either wave, and so AW propagation is entirely unaffected. The only modification is to the ratio of the real and complex parts of the IAW frequency $\omega_{\rm C,i}/\omega_{\rm C,r}$, which, unless the electrons partake in Landau damping of the acoustic mode, decreases with increasing $T_{\rm e}$. This would bolster the effectiveness of the IAW--AW interaction by extending the lifetime of the pressure anisotropy and allowing the interaction to occur over more than one linear time of the frequency-matched modes. 

{\color{black} In \S\ref{sec:form}, we noted the lack of $k_\perp$ as a key difference between these pressure-anisotropy-induced interactions and those found within standard Alfv\'enic turbulence. The nature of the pressure-anisotropic stress we highlight is significantly different from the Reynolds and Maxwell stresses, which dominate energy transfer in Alfv\'enic turbulence \citep{grete17}. Most importantly, $\beta\Delta$, being a scalar, will always affect $v_{\rm A,eff}$ regardless of the ratio $k_\perp/k_\|$, doing so through the same nonlinearity that we retain in \eqref{eq:elsass}. Therefore, the effects of IAWs or AWs with $\beta\Delta\sim 1$ are expected to remain leading order when $k_\perp \neq 0$. Furthermore, the linear timescales of IAWs and AWs are independent of $k_\perp$, thus the frequency matching criterion of their interaction, and its spectral non-locality, are unlikely to change.}

Given the further complications of {\color{black} microinstabilities} and other high-$\beta$ effects such as magneto-immutability \citep{squire23}, formulating an analytical model for turbulence {\color{black}in this parameter regime} remains a challenging task. {\color{black} The intention of this work is primarily to hint at the physics underlying changes in the observed behaviour of collisionless turbulence at $\beta\gg 1$, such as non-local energy transfer due to the pressure-anisotropic stress \citep{arzamasskiy23}}. {\color{black}To that end}, we may still speculate upon some of the more qualitative consequences of the wave--wave interactions described herein. One of the most effective tools for describing strong turbulence is an appropriately reduced set of equations. For example, reduced MHD incorporates the concept of critical balance directly by ordering the linear frequency comparable to the nonlinear frequency, $\tau_{\rm lin}^{-1} \doteq k_\parallel v_{\rm A} \sim \tau_{\rm nl}^{-1} \approx k_\perp u_\perp$ \citep{gs95,schekochihin09}. Reinstating nonzero $k_\perp$ then, we can consider the reduced Alfv\'enic Els\"asser equations that result from our 1D high-$\beta$ ordering \eqref{eq:ord}, modified to include $k_\|/k_\perp \sim \delta B_\parallel/B_0 \sim \epsilon$:
\begin{multline}\label{eq:2delsass}
    \pD{t}{\bb{z}^\pm_\perp} \mp v_{\rm A} \nabla_\parallel \bb{z}^\pm_\perp + \bb{z}^\mp_\perp\bcdot\grad_\perp \bb{z}^\pm_\perp + \grad_\perp P_{\rm total} \\*
    = v_{\rm A}\frac{\beta}{4}\nabla_\parallel\bigl[(\bb{z}_\perp^+-\bb{z}_\perp^-)\Delta\bigr]
    + \frac{\beta}{4}(\bb{z}_\perp^+-\bb{z}_\perp^-)\bcdot\grad_\perp\bigl[(\bb{z}_\perp^+-\bb{z}_\perp^-)\Delta\bigr],
\end{multline}
where $P_{\rm total}$ represents the combined perpendicular and magnetic pressures. Contained within the right-hand side of \eqref{eq:2delsass} are the various new effects discussed throughout this work that are absent in a collisional MHD model, or even a $\beta\sim \mathcal{O}(1)$ gyrokinetic model. The first term on the right-hand side incorporates the ability of AWs to travel at different $v_{\rm A,eff}$, while both terms describe the deformation in $\bb{z}^\pm$ that results from wave--wave interactions. Equation~\eqref{eq:2delsass} does not distinguish between the source of $\beta \Delta$; be it generated by AWs with $\delta B_\perp/B_0 \sim \beta^{-1/2}$ or IAWs with $\delta\rho/\rho \sim \beta^{-1}$, the Alfv\'enic cascade will be modified. In each case, however, the description of the source of fluctuations can be more difficult. For example, IAWs act non-locally to affect Alfv\'enic fluctuations in $k$-space, but the reverse may not be true for Alfv\'enic mixing of acoustic waves. In that case, there is not necessarily a single acoustic Els\"{a}sser equation that can be written at each scale, because multiple $\boldsymbol{k}_\perp$ and $k_\parallel$ are inherent to the problem. Furthermore, in a collisionless turbulent cascade, $\beta\Delta$ cannot remain $\mathcal{O}(1)$ at every wavenumber throughout the inertial range. As all of the other quantities cascade, $\beta$ will remain constant, so it is likely that interactions mediated by $\Delta$ are a leading-order effect only in a specific portion of the inertial range where $\beta \Delta \sim 1$. One subtle conclusion of the above reduced equations is that pressure balance in these plasmas should be struck primarily between $B^2$ and the perpendicular pressure $p_\perp$, rather than the isotropic pressure \citep{squire23}. The leading order of \eqref{eq:2delsass} implies that $\delta p_\perp/p_0 \sim \beta^{-1}\delta B_\parallel/B_0 \sim \epsilon^2$. In essence, the anisotropy generated should be dominated by the parallel pressure perturbation. This was found empirically in simulations of incompressibly driven turbulence using the same CGL-MHD \texttt{Athena++} code employed here \citep{squire23}, as well as in the hybrid-kinetic simulations of \citet{arzamasskiy23}.

Additional effects may be uncovered by studying the weak turbulence resulting from these wave--wave interactions. Isolated wave packets are not typically considered within strong turbulence, however the long deformation times associated with weak turbulence allows waves to remain correlated over many local Alfv\'en-crossing times \citep{iroshnikov64,kraichnan68}. This also expands upon the range of scales over which $\beta \Delta$ exerts an influence by not requiring that the interaction remain strong. One effect that may be impactful to weak turbulence is the ability of AWs to modify each others' $k_\parallel$. Current understandings of weak AW interactions involve resonance conditions describing three-wave interactions \citep{galtier2000}:
\begin{align}
    \boldsymbol{k}_1 + \boldsymbol{k}_2 &= \boldsymbol{k}_3 &\Longrightarrow&  &k_{\parallel1}+k_{\parallel2}&=k_{\parallel3} \\
    \omega(\boldsymbol{k}_1^\pm) + \omega(\boldsymbol{k}_2^\mp) &= \omega(\boldsymbol{k}_3^\pm) &\Longrightarrow& & k_{\parallel1}-k_{\parallel2}&=k_{\parallel3},
\end{align}
which demand that $k_{\parallel 2}=0$ and $k_{\parallel 1} = k_{\parallel 3}$. However, these resonance conditions assume that $v_{\rm A}(k_\parallel) = v_{\rm A}$, which is subject to modification if $v_{\rm A,eff}$ is a function of the fluctuation amplitude, and hence the scale $k_\parallel$ of concern. In fact, for $\beta \Delta \ll 1$, the frequency resonance condition would become
\begin{equation}
    v_{\rm A}k_{\parallel 1}\biggl[1+  \frac{\beta}{4} \Delta(k_{\parallel 1})\biggr] - v_{\rm A}k_{\parallel 2}\biggl[1 + \frac{\beta}{4} \Delta(k_{\parallel 2})\biggr] \approx v_{\rm A}k_{\parallel 3} \biggl[1+\frac{\beta}{4} \Delta(k_{\parallel 3})\biggr].
\end{equation}
Unlike in weak MHD Alfv\'enic turbulence, a $k_{\parallel 2} = 0$ mode is no longer a solution of the above condition by virtue of $\Delta(k_{\parallel 2})$ \textit{increasing} with decreasing $k_{\parallel 2}$ (assuming a cascade to higher wavenumbers for $\Delta$). This holds regardless of the additional fact that co-propagating interactions can occur, which have been shown to change $k_\parallel$ as well. Together these effects may be capable of modifying the onset of critical balance, changing the transition scale between weak and strong turbulence. Well-crafted simulations of Alfv\'enic and compressive turbulence in both collisionless and weakly collisional high-$\beta$ plasmas are the clear next step.

%
%
\section*{Acknowledgements}

The authors thank Jonathan Squire, Alexander Schekochihin, Eliot Quataert, and the participants of the 14th Plasma Kinetics Working Meeting at the Wolfgang Pauli Institute in Vienna for useful discussions, {\color{black} as well as the two anonymous reviewers for comments that helped sharpen the presentation.}

\section*{Funding}

S.M.~and M.W.K.~were supported in part by NSF CAREER Award No.~1944972. High-performance computing resources were provided by the PICSciE-OIT TIGRESS High Performance Computing Center and Visualization Laboratory at Princeton University.

\section*{Declaration of Interests}

The authors report no conflict of interest.

%
%
\appendix

\section{Energy conservation in IAW--AW interactions}\label{sec:appecons}

Without going to higher order in $\epsilon$ when deriving an equation for the evolution of the ion acoustic wave, the solution \eqref{eq:collsln} describing the interaction of AWs with ion acoustic waves does not conserve the total energy. The reasoning for this is that, while the effect of the acoustic wave's pressure anisotropy on the AW is leading order in $\epsilon$, the back reaction of the AW on the acoustic wave occurs at $\mathcal{O}(\epsilon^2)$. As assurance for the concerned reader, however, we demonstrate here that by going to at least third order in the compressive dynamical equations, energy conservation can be reestablished. Additionally, this result aids in confirming that the frequency matching condition $k_{\rm A}  \sim \sqrt{\beta}k_{\rm C}$ does not violate the assumptions used to obtain the ordered equations \eqref{eq:elsass} and \eqref{eq:lincomp}. To achieve this task, we begin by writing down the total energy of the background plus perturbations, which we normalize using $\rho_0 v_{\rm th,i}^2 = 2p_0$ to make the orders of each contribution more apparent. Following this we evaluate the time derivative of the total energy using the evolution equations \eqref{eq:fullcgl}, with the proper adjustments made for our assumptions about the problem geometry. Once this has been done, we examine the orders of the remaining terms to demonstrate that energy conservation is reestablished by including all terms of order $\epsilon^3$ or higher.

The total energy normalized to $\rho_0 v_{\rm th,i}^2$ is given by
\begin{equation}\label{eq:app1}
    \frac{E}{\rho_0 v_{\rm th,i}^2} = \int \mathrm{d}x\, \biggl\{  
    \frac{1}{\beta}\frac{\rho}{\rho_0} \biggl( \frac{u_\perp}{v_{\rm A}} \biggr)^2 
    +\frac{1}{\beta}\biggl(1+\frac{\delta B_\perp^2}{B_0^2} \biggr) + \frac{\rho}{\rho_0} \biggl( \frac{u_\parallel}{v_{\rm th,i}}\biggr)^2 + \frac{p_\perp}{p_0}+ \frac{1}{2}\frac{p_\parallel}{p_0}
    \biggr\}.
\end{equation}
Already, it is clear from \eqref{eq:app1} that, according to the ordering \eqref{eq:ord}, the Alfv\'enic contributions to the energy (terms involving $u_\perp$ and $\delta B_\perp$) are at most of order $\epsilon^3$, while the compressive contributions (the last three terms) are up to linear in $\epsilon$. Taking the time derivative of \eqref{eq:app1}, the rate of change of the total energy is given by
\begin{multline}\label{eq:dedt}
    \frac{1}{\rho_0 v_{\rm th,i}^2}\frac{\mathrm{d} E}{\mathrm{d}t} = \int \mathrm{d}x\, \biggl\{  
    \frac{1}{\beta} \pD{t}{} \biggl(\frac{\delta \rho}{\rho_0}\biggr) \biggl( \frac{u_\perp}{v_{\rm A}} \biggr)^2 +
    \frac{1}{\beta} \biggl( \frac{\rho}{\rho_0} \biggr)  \pD{t}{} \biggl( \frac{u_\perp}{v_{\rm A}} \biggr)^2 + \frac{1}{\beta} \pD{t}{} \biggl( \frac{\delta B_\perp}{B_0} \biggr)^2 \\
    + \pD{t}{} \biggl(\frac{\delta \rho}{\rho_0}\biggr) \biggl( \frac{u_\parallel}{v_{\rm th,i}} \biggr)^2 +
     \biggl( \frac{\rho}{\rho_0} \biggr)  \pD{t}{} \biggl( \frac{u_\parallel}{v_{\rm th,i}} \biggr)^2  + \pD{t}{}\frac{\delta p_\perp}{p_0} +\frac{1}{2}\pD{t}{}\frac{\delta p_\parallel}{p_0}
    \biggr\}.
\end{multline}
We examine these terms in turn with the aim of establishing the lowest order at which they appear.

First, we focus on those terms relating directly to the energy carried by the Alfv\'enic fluctuations, \textit{i.e.}, the top line of \eqref{eq:dedt}. Denote the characteristic Alfv\'enic and compressive wavenumbers by $k_{\rm A}$ and $k_{\rm C}$, respectively. The first term is of order $k_{\rm C} \vcs \epsilon^4$ (due to the continuity equation's compressive nature in $\partial_t$), while the second and third terms are of order $k_{\rm A} \valf \epsilon^3$. For all circumstances studied in this paper then (including frequency-matched fluctuations), the leading order of the second and third terms will dominate and be the main mechanism by which the Alfv\'enic fluctuations extract energy from the compressive ones. Therefore, neglecting the first term and inserting \eqref{eq:redalf} for $\partial_t u_\perp$ and $\partial_t \delta B_\perp$, the energy rate of change simplifies to
\begin{multline}\label{eq:nrg_redalf}
    \frac{1}{\rho_0 v_{\rm th,i}^2}\frac{\mathrm{d} E}{\mathrm{d}t} \approx \int \mathrm{d}x \,\biggl\{  
    u_\perp \pD{x}{}\biggl(\frac{\delta B_\perp}{B_0} \frac{\Delta p}{p_0} \biggr)\\
    + \pD{t}{} \biggl(\frac{\delta \rho}{\rho_0}\biggr) \biggl( \frac{u_\parallel}{v_{\rm th,i}} \biggr)^2 +
     \biggl( \frac{2\rho}{\rho_0} \biggr) \frac{u_\parallel}{v_{\rm th,i}}  \pD{t}{} \biggl( \frac{u_\parallel}{v_{\rm th,i}} \biggr) + \pD{t}{}\frac{\delta p_\perp}{p_0} + \frac{1}{2}\pD{t}{}\frac{\delta p_\parallel}{p_0}
    \biggr\}.
\end{multline}
For $\omega_{\rm A} \approx \omega_{\rm C}$, the remaining Alfv\'enic term on the first line of \eqref{eq:nrg_redalf} is of the exact same order as the kinetic energy terms on the bottom line, while, for $k_{\rm A} \approx k_{\rm C}$, it is one half-order in $\epsilon$ smaller. In order to satisfy energy conservation, we must then include all compressive terms that are up to, but \textit{not} including, those of order $\epsilon^4$. 

Before embarking on the compressive terms however, we note that the heat fluxes need not be included in \eqref{eq:cglpprl} and \eqref{eq:cglpprp} for the proof of energy conservation between Alfv\'en and acoustic waves. Their effect is to diffuse wave energy into the background, and not to facilitate communication between the Alfv\'enic and compressive fluctuations. We therefore proceed by evolving the pressure perturbations according to the double adiabats \eqref{eq:da}, which can be combined in the following manner:
\begin{multline}\label{eq:daord}
    \pD{t}{p_\perp} + \frac{1}{2} \pD{t}{p_\parallel} = \frac{1}{2}\frac{\Delta p}{B^2}\pD{t}{B^2} + \frac{3p_\parallel}{2\rho}\pD{t}{\rho}  - \frac{u_\parallel}{2}\pD{x}{p_\parallel} + \frac{3u_\parallel}{2} \biggl( \frac{p_\parallel}{\rho}\biggr) \pD{x}{\rho}\\
    + \frac{p_\perp}{\rho} \pD{t}{\rho} - u_\parallel\rho\pD{x}{}\biggl(\frac{p_\perp}{\rho}\biggr) 
     + u_\parallel \biggl(\frac{p_\perp}{B}\biggr) \pD{x}{B} - u_\parallel\biggl(\frac{p_\parallel}{B}\biggr) \pD{x}{B} .
\end{multline}
Noting that $\delta B \propto \delta B_\perp^2$, which makes $B^{-1}\partial_xB$ second order in $\epsilon$, the final two terms cancel up to and including $\mathcal{O}(\epsilon^3)$. We also use the fact that \eqref{eq:daord} will be integrated over space to swap the remaining $x$-derivative on the bottom line from $p_\perp/\rho$ to $u_\parallel \rho$, picking up a sign change. Using the continuity equation to cancel the first two terms on the second line, no more of the second line remains. Normalization and integration of the first term on the right-hand side of \eqref{eq:daord} then gives
\begin{equation}
    \int \mathrm{d}x\, \frac{1}{2} \biggl(\frac{\Delta p}{p_0}\biggr) \frac{\delta B_\perp}{B_0} \pD{t}{} \biggl(\frac{\delta B_\perp}{B_0}\biggr) = - \int \mathrm{d}x \,\frac{u_\perp}{2}\pD{x}{}\biggl(\frac{\delta B_\perp}{B_0} \frac{\Delta p}{p_0} \biggr),
\end{equation}
where \eqref{eq:redalfind} has been used to substitute for $\partial_t \delta B_\perp$. This term will cancel the first term on the right-hand side of \eqref{eq:nrg_redalf}, leaving us with the following expression for the energy rate of change:
\begin{multline}\label{eq:nrg_comp}
    \frac{1}{\rho_0 v_{\rm th,i}^2}\frac{\mathrm{d} E}{\mathrm{d}t} \approx \int \mathrm{d}x \,\biggl\{  
     \frac{3p_\parallel}{2p_0}\biggl(\frac{1}{\rho} \pD{t}{\rho}\biggr)  - \frac{u_\parallel}{2p_0}\pD{x}{p_\parallel} + \frac{3u_\parallel}{2\rho} \biggl( \frac{p_\parallel}{p_0}\biggr) \pD{x}{\rho} \\*
     \mbox{} + \pD{t}{} \biggl(\frac{\delta \rho}{\rho_0}\biggr) \biggl( \frac{u_\parallel}{v_{\rm th,i}} \biggr)^2 +
    \biggl( \frac{2\rho}{\rho_0} \biggr) \frac{u_\parallel}{v_{\rm th,i}}  \pD{t}{} \biggl( \frac{u_\parallel}{v_{\rm th,i}} \biggr) \biggr\} + \mathcal{O}(\epsilon^4).
\end{multline}
At this point, it is may become obvious that, with all of the remaining terms being compressive in nature, they must cancel because the nonlinear interaction of a wave with itself generally conserves energy in the absence of dissipative effects like heat fluxes or collisions. Regardless, to solidify the argument we will continue and demonstrate that there are no other Alfv\'enic feedback terms that appear at orders greater than $\epsilon^4$. Using the continuity equation to simplify the top line of \eqref{eq:nrg_comp} yields
\begin{equation}
  \int \mathrm{d}x \,\biggl\{ - \frac{3p_\parallel}{2p_0}\biggl[\frac{1}{\rho} \pD{x}{} (\rho u_\parallel)\biggr]  - \frac{u_\parallel}{2p_0}\pD{x}{p_\parallel} + \frac{3u_\parallel}{2\rho} \biggl( \frac{p_\parallel}{p_0}\biggr) \pD{x}{\rho} \biggr\} = \int \mathrm{d}x\,  \frac{u_\parallel}{p_0}\pD{x}{p_\parallel},
\end{equation}
which, after once again employing continuity, reduces \eqref{eq:nrg_comp} to 
\begin{equation}\label{eq:simp_nrgcomp}
    \frac{1}{\rho_0 v_{\rm th,i}^2}\frac{\mathrm{d} E}{\mathrm{d}t} \approx \int \mathrm{d}x\, \biggl\{  
      \frac{u_\parallel}{p_0}\pD{x}{p_\parallel} +
    \frac{2\rho}{\rho_0} \biggl( \frac{u_\parallel}{v_{\rm th,i}} \biggr)^2 \pD{x}{u_\parallel} +
    \biggl( \frac{2\rho}{\rho_0} \biggr) \frac{u_\parallel}{v_{\rm th,i}}  \pD{t}{} \biggl( \frac{u_\parallel}{v_{\rm th,i}} \biggr) \biggr\} + \mathcal{O}(\epsilon^4).
\end{equation}
The fully nonlinear parallel momentum in this case is given by 
\begin{equation}\label{eq:parmom}
    \rho \pD{t}{u_\parallel} = - \rho u_\parallel \pD{x}{u_\parallel} - \pD{x}{p_\parallel} - \pD{x}{} \biggl(\frac{\delta B_\perp^2}{B^2}\Delta p \biggr) - \delta B_\perp \pD{x}{\delta B_\perp} .
\end{equation}
Multiplying \eqref{eq:parmom} by $u_\parallel/\rho_0 \vth^2 = u_\parallel/2p_0$, the last two terms on its right-hand side become of order $\mathcal{O}(\epsilon^4)$, while the first two cancel the remaining terms in \eqref{eq:simp_nrgcomp}. As a result, we have shown that, up to and including $\mathcal{O}(\epsilon^3)$ terms, total energy is conserved by the interaction of Alfv\'en and ion-acoustic waves according to the ordering \eqref{eq:ord}. This finding tells us that, even when the frequencies of the two modes are matched and $k_{\rm A} \sim \sqrt{\beta}k_{\rm C}$, the leading order for the dynamical equations describing each wave remains unchanged. Naturally, no terms neglected in deriving \eqref{eq:redalf} are enhanced with respect to the leading order by letting $k_{\rm A}$ increase by $\epsilon^{-1/2}$. At the same time, Alfv\'enic feedback effects on the compressive fluctuations were of order $\epsilon^3$, meaning the leading order compressive equations \eqref{eq:lincomp} remain the same until $\omega_{\rm A} \sim \beta^{2}\omega_{\rm C}$. Nonlinear steepening will remain far more important in the frequency-matched regime of strong interaction, thus for (small) acoustic fluctuations of amplitude ${\sim}\beta^{-1}$, the assumption of linearity is quite robust.

%
%
\section{The $\Delta$ profile of a Landau-fluid AW packet}\label{sec:appawdp}

This appendix provides a derivation of the pressure anisotropy generated by an isolated, parallel-propagating, large-amplitude ($\delta B_\perp/B_0 \sim \beta^{-1/2}$) AW packet, specifically for the diffusive heat fluxes of a Landau-fluid-type model. As discussed in \S\ref{sec:awaw}, we predict that AWs of this type will steepen to form shocks in the profile of their $\delta B_\perp$ and $u_\perp$ perturbations. These shocks will propagate at a speed near $v_{\rm A}$, which is much slower than the thermal speed of the plasma, allowing the change in pressure anisotropy generated by $\delta B_\perp$ to diffuse upstream ahead of the shock. Here we derive the functional form of $\Delta(x)$ as it propagates with the shock. 

The first step is to introduce another ordering that describes this large-amplitude AW near the interruption limit and which is distinct from \eqref{eq:ord}. The need for a new ordering originates from the fact that in the AW-AW interaction, the primary AW cannot be derived from the ordering \eqref{eq:ord} as \eqref{eq:ord} relies on linear perturbations to produce the pressure anisotropy. As stated in \S\ref{sec:awaw}, the primary AW induces pressure anisotropy nonlinearly. Therefore in order to produce $\Delta \sim \beta^{-1}$, it must have $\delta B_\perp/B_0$ larger than that of the secondary packet, which obeys the Alfv\'enic components of \eqref{eq:ord}. Our new ordering must then capture the fact that AWs perturb the pressure and other compressive fields nonlinearly, while being consistent with the high-$\beta$ and $\beta\Delta\sim 1$ assumptions of \eqref{eq:ord} (such that the resultant $\Delta(x)$ may be used in \eqref{eq:elsass}):
\begin{equation}\label{eq:aword}
    \frac{\delta B_\perp^2}{B_0^2} \sim \frac{u_\perp^2}{v_{\rm A}^2} \sim \frac{\delta \rho}{\rho_0} \sim \frac{u_\parallel}{v_{\rm A}} \sim \frac{1}{\beta} \sim \Delta \sim \sigma^2 \sim \epsilon.
\end{equation}
Here, $\sigma$ is a new expansion parameter satisfying $\sigma \gg \epsilon$. The estimates of $\partial_t$ for compressive and Alfv\'enic fields are the same as those used with \eqref{eq:ord} (${\sim}k \vcs$ and ${\sim}kv_{\rm A}$, respectively). Note that as an ordering, this is unaffected by the Alfv\'enic dynamics following $v_{\rm A,eff}$ rather than $v_{\rm A}$, as the pressure anisotropy is strictly positive and thus will never approach the firehose limit where $v_{\rm A,eff} \ll v_{\rm A}$.

\begin{figure}
    \centering
    \includegraphics[width=0.97\textwidth]{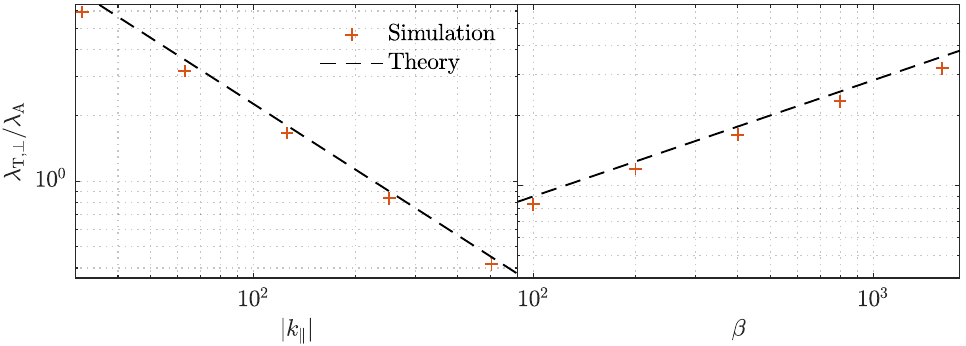}
    \caption{Comparison of the analytically predicted and simulated decay lengths of the pressure anisotropy precursor of a steepened AW packet, for $\bb{q}_{\perp/\|}$ given by the `3+1' heat fluxes \eqref{eq:fullhf}. The calculated decay length $\lambda_{\rm T,\perp}$ is normalized by the dominant wavelength of the AW packet $\lambda_{\rm A}$. On the left, the decay length is varied with respect to the Landau wavenumber $|k_\||$, while on the right it is varied with respect to $\beta$.}
    \label{fig:AWtheorycomp}
\end{figure}

The next step is to use \eqref{eq:aword} to obtain an estimate for the size of the parallel pressure perturbation $\delta p_\parallel$, and show that it is much smaller than the perpendicular pressure perturbation $\delta p_\perp$. We apply \eqref{eq:aword} to the parallel momentum equation evaluated downstream of the shock (where the gradient in $\delta B_\perp$ is nonexistent), finding that the leading-order equation is simply linear:
\begin{equation}
    \rho_0 \frac{\partial u_\parallel}{\partial t} \approx -\frac{\partial \delta p_\parallel}{\partial x}.
\end{equation}
This approximate equation implies that $\delta p_\parallel \sim \sigma^2 \rho_0 v_{\rm A}\vcs$, or $\delta p_\parallel/p_0 \sim \sigma^3$. Comparatively, no such restraint is placed on $\delta p_\perp$ by the perpendicular momentum equation, so the pressure anisotropy must be dominated by $\delta p_\perp$. The equation for $\delta p_\perp$ is in this case more easily formulated in terms of the perpendicular temperature $\delta T_\perp$, which according to \eqref{eq:aword} and \eqref{eq:cglpprp} evaluated downstream from the shock satisfies
\begin{equation}\label{eq:appdiff}
    \frac{\partial \delta T_\perp}{\partial t} \approx \kappa \frac{\partial ^2 \delta T_\perp}{\partial x^2},
\end{equation}
where $\kappa$ is a diffusion coefficient (equal to $\rho_0v_{\rm th}/\sqrt{\pi}|k_\||$ in the `3+1' Landau-fluid model; \citealt{shd97}). Given that $\kappa$ operates on thermal timescales, the diffusion of $\delta T_\perp$, and thus $\Delta p$, downstream is effectively instantaneous as the AW packet propagates. In the frame of the shock then, $\delta T_\perp$ attains a steady-state profile, meaning that it can be represented as $\delta T_\perp (x-v_{\rm A,max}t)$ (where $v_{\rm A,max}$ is a constant equal to the peak value of $v_{\rm A,eff}$ evaluated at the shock front). Accordingly, the relation \eqref{eq:appdiff} becomes
\begin{equation}\label{eq:diffsln}
    \frac{\partial^2 \delta T_\perp}{\partial x^2} + \frac{v_{\rm A,max}}{\kappa} \frac{\partial \delta T_\perp}{\partial x} \approx 0 \quad \Longrightarrow \quad \delta T_\perp \approx \delta T_{\rm \perp,max} \exp \biggr[\frac{v_{\rm A,max}}{\kappa}(v_{\rm A,max}t-x)\biggr].
\end{equation}
This gives the functional form of $\delta T_\perp$ leading up to the shock front (where $\delta T_{\perp, \rm max}$ is determined), and motivates our choice of exponentially decaying pressure anisotropy as a source for \eqref{eq:elsass} used in \S\ref{sec:awaw}. To test the accuracy of this estimate of the decay length of $\delta T_\perp$, a series of Landau-fluid simulations were performed with varying choices of $\beta$ and for $|k_\||$ in \eqref{eq:fullhf}, shown in figure \ref{fig:AWtheorycomp}. The setup of the wave packet in these runs is the same as that of the primary packet described in \S\ref{sec:numawaw}. The decay length of $\delta T_\perp$ is found in these simulations by waiting until the wave packet steepens to form a shock, then fitting an exponentially decaying function to the region directly in front of the peak of the magnetic-field perturbation. Compared with the theoretical estimate using the `3+1' heat fluxes, strong agreement is found.

The actual magnitude of $\delta T_{\perp, \rm max}$ is determined by acknowledging that $\partial_x \delta B_\perp^2$ is large in the region of the shock, and including its associated term at leading order in the equation left of the arrow in \eqref{eq:diffsln} provides
\begin{equation}
    \frac{\partial^2 \delta T_\perp}{\partial x^2} + \frac{v_{\rm A,max}}{\kappa} \frac{\partial \delta T_\perp}{\partial x}  \approx \frac{v_{\rm A,max} T_0}{2\kappa B_0^2}\frac{\partial \delta B_\perp^2}{\partial x}.
\end{equation}
Integrating this equation from the shock front to the distant downstream region where $\partial_x \delta T_\perp \approx 0$, we obtain the simple result that $\delta T_{\perp, \rm max}/T_0 \approx (1/2)\delta B_{\perp, \rm max}^2/B_0^2$. Knowing that $\delta p_\parallel$ may be neglected, the same steps can be applied to \eqref{eq:cglpprl} to obtain a diffusion equation for $\delta \rho$, which is found to decay on a length scale of $(3/2)\kappa/v_{\rm A,max}$ instead of  $\kappa/v_{\rm A,max}$. Similarly, $\delta \rho_{\rm max}$ is determined to be $\delta \rho_{\rm max}\approx (1/3)\delta B_{\perp, \rm max}^2/B_0^2$. Combined, these yield the final result that $\Delta_{\rm max} \approx \delta p_{\perp, \rm max}/p_0 \approx (5/6)\delta B_{\perp, \rm max}^2/B_0^2$, and $v_{\rm A,max} = v_{\rm A}\sqrt{1+\beta\Delta_{\rm max}/2}$.


\end{document}